\documentclass[varenna]{cimento}

\usepackage[utf8]{inputenc}
\usepackage[T1]{fontenc}
\usepackage{graphicx}%
\usepackage{multirow}%
\usepackage{amsmath,amssymb,amsfonts}%
\usepackage{amsthm}%
\usepackage{mathrsfs}%
\usepackage[title]{appendix}%
\usepackage{xcolor}%
\usepackage{textcomp}%
\usepackage{manyfoot}%
\usepackage{booktabs}%
\usepackage{algorithm}%
\usepackage{algorithmicx}%
\usepackage{algpseudocode}%
\usepackage{listings}%
\usepackage{cancel}%
\usepackage{bm}
\usepackage{mathtools}
\usepackage{orcidlink}

\newcommand{\beq}{\begin{equation}}
\newcommand{\eeq}{\end{equation}}
\newcommand{\bea}{\begin{eqnarray}}
\newcommand{\eea}{\end{eqnarray}}

\makeatletter
\renewenvironment{thebibliography}[1]{%
  \par\bigskip
  \noindent{\bfseries \MakeUppercase\refname}\par\smallskip
  \begin{marginlist}[\usecounter{enumiv}%
    \let\p@enumiv\@empty
    \renewcommand{\theenumiv}{\arabic{enumiv}}]%
    {\@biblabel{\arabic{enumiv}}}{\@biblabel{#1}}%
  \sloppy\hyphenpenalty10000\clubpenalty4000\widowpenalty4000%
  \sfcode`\.=\@m \small
}{%
  \end{marginlist}%
}
\makeatother

\begin{document}

\title{Superfluidity in Fermi systems within the framework of Density Functional Theory}
\author{Piotr Magierski~\orcidlink{0000-0001-8769-5017} } 
\institute{Faculty of Physics, Warsaw University of Technology, \\
ulica Koszykowa 75, Warsaw, 00-662, Poland}
\institute{Physics Department, University of Washington, \\ 3910 15th Ave. NE, Seattle, 
WA 98195-1560, USA} 

\date{June 2024}

\maketitle

\begin{abstract}
This review is based on lectures given by the author at the 
{\em Enrico Fermi} Summer School in Varenna.
It presents the basics of Density Functional Theory (DFT) for Fermi superfluids, with particular emphasis on nuclear systems. Special attention is given to the foundations of both DFT and time-dependent DFT (TDDFT). The review explores the advantages and challenges involved in the practical application of TDDFT to superfluid systems, as well as the typical approximations employed. Various applications of the TDDFT framework to the description of phenomena related to nonequilibrium superfluidity in atomic nuclei, neutron stars, and ultracold atoms are discussed.
\end{abstract}

\section{Introduction}

Density Functional Theory (DFT) has become nowadays a standard theoretical tool for studies of interacting many-body Fermi systems~\cite{RevModPhys.71.1253,dreizler2012density,eschrig1996fundamentals,parr1989density}.
It offers a universal and formally exact approach, which had enormous practical successes. 
In the situation, where the properties of a system of large number of strongly interacting fermions need to be found, it is difficult to find a better theoretical method, with the same versatility in tackling
variety of problems associated with the description of many-body systems. It provides the framework,
which is computationally tractable and avoids various simplifying assumptions.
One may argue, that the methods known under the general name of Quantum Monte Carlo, which aim at solving the problem "exactly" starting from interparticle interactions offer a more reliable, and controllable approach (see e.g., \cite{RevModPhys.87.1067} for a review).
However, they are limited to stationary problems and due to numerical complexity it is practically impossible to
use them in the case of large and inhomogeneous systems.
Moreover, if the nonstationary problems are considered, especially those where the system is far from equilibrium,
there is no real competitor of density functional theory.
The machinery of nonequilibrium Greens function approach, especially Keldysh method of treating nonequilibrium problems cannot compete with Density Functional Theory approach. They may offer, however,
a better insight into the structure of energy density functionals for the time dependent problems (see Ref. \cite{RevModPhys.58.323} for a review of Keldysh formalism and Ref. \cite{vanleeuwen2005introductionkeldyshformalismapplications} - for its application
to TDDFT).

There is however a significant difference  between DFT
and other theoretical tools of quantum many-body physics.
The central object in DFT is the energy density functional which 
is merely proved to exist by means of the Hohenberg-Kohn theorem~\cite{PhysRev.136.B864}.
In the case of low energy nuclear dynamics, in particular when
non-magic medium or heavy nuclei are involved, the proper treatment of superfluidity is crucial 
and the conventional DFT descripiton has to be extended.
The first attempt to develop the formal framework of DFT for superconductors has been triggered
by the discovery of high-temperature superconductivity~\cite{PhysRevLett.60.2430}.
In this review, I will walk the reader through Density Functional Theory (DFT) and its description of many-body interacting systems, with a particular focus on its extension to fermionic superfluids. I will present the main results concerning the foundations of DFT and its time-dependent formulation (TDDFT). Specifically, I will discuss the most useful implementation of TDDFT—the superfluid local density approximation—which provides a powerful theoretical framework for studying nonequilibrium superfluid dynamics in Fermi systems.

\section{Hohenberg-Kohn theorem}

The main idea of DFT is to replace the many-body wave function, which is enormously complicated
object, by the density $\rho(\bf r)$, which is clearly much simpler quantity. This approach was considered even before DFT has been formulated, 
however, as an approximate description of a many body system.
Indeed, let us consider the Hamiltonian 
describing spin-1/2 fermions of mass $m$ interacting via two body interaction
$V^{int}_{\sigma,\sigma'}(\bm{r},\bm{r'})$:
\begin{eqnarray}\label{Hamiltonian}
\hat{H}&=&\hat{T} + \hat{V}^{ext} + \hat{V}^{int} =
\sum_{\sigma=\uparrow,\downarrow}\int d^{3}\bm{r}\;
\hat{\psi}_{\sigma}^\dagger(\bm{r})
\left[-\frac{\hbar^2}{2m}\nabla^2 + V_{\sigma}^{ext}(\bm{r})\right]
\hat{\psi}_\sigma(\bm{r}) \nonumber \\
&+& \frac{1}{2}\sum_{\sigma,\sigma' =\uparrow,\downarrow}
\int d^{3}\bm{r}\int d^{3}\bm{r'}
\hat{\psi}_{ \sigma}^\dagger\left(\bm{r}\right)
\hat{\psi}_{\sigma'}^\dagger\left(\bm{r'}\right) V^{int}_{\sigma,\sigma'}(\bm{r},\bm{r'})
\hat{\psi}_{\sigma'}        \left(\bm{r'}\right)
\hat{\psi}_{ \sigma}        \left(\bm{r}\right),
\end{eqnarray}
where $V_{\sigma}^{ext}(\bm{r})$ is the external potential ($\sigma, \sigma'$ denote spin projection indices).
Then the prototype of density functional can be constructed by considering $\langle \hat{H} \rangle$
and using Hartree approximation~\cite{Hartree_1928} to express the interaction term in the form:
\begin{eqnarray}
\langle\hat{V}\rangle = \frac{1}{2}\sum_{\sigma,\sigma' =\uparrow,\downarrow}
\int d^{3}\bm{r}\int d^{3}\bm{r'}
V^{int}_{\sigma,\sigma'}(\bm{r},\bm{r'})\rho_{\sigma}({\bf r})\rho_{\sigma'}({\bf r'}),
\end{eqnarray}
where $\rho_{\sigma}({\bf r}) = \langle \hat{\psi}_{ \sigma}^\dagger\left(\bm{r}\right) \hat{\psi}_{ \sigma}\left(\bm{r}\right)\rangle$ is the density distribution of fermions with spin-projection $\sigma$. On the other hand the kinetic term can be approximated by:
\begin{eqnarray}
\langle\hat{T}\rangle = \frac{3}{5}\frac{\hbar^2}{2m}(6\pi^{2})^{2/3}\sum_{\sigma =\uparrow,\downarrow}
\int d^{3}\bm{r}\left [ \rho_{\sigma}({\bf r}) \right ]^{5/3},
\end{eqnarray}
which is known as Thomas-Fermi approximation~\cite{Thomas_1927,Fermi_1927}. As a result one gets an
approximate density functional defined as:
\begin{eqnarray}\label{thomasfermi}
E_{TF}[\rho ] = \sum_{\sigma =\uparrow,\downarrow}
\int d^{3}\bm{r}\left (\frac{3}{5}\frac{\hbar^2}{2m}(6\pi^{2})^{2/3}
\left [ \rho_{\sigma}({\bf r}) \right ]^{5/3} + 
V_{\sigma}^{ext}(\bm{r})\rho_{\sigma}({\bf r}) \right ) + \nonumber \\ + \frac{1}{2}\sum_{\sigma,\sigma' =\uparrow,\downarrow}
\int d^{3}\bm{r}\int d^{3}\bm{r'}
V^{int}_{\sigma,\sigma'}(\bm{r},\bm{r'})\rho_{\sigma}({\bf r})\rho_{\sigma'}({\bf r'}).
\end{eqnarray}
This functional is usually used as a first approximation to describe
the ground state of many-body system. The density corresponding to the ground
state can be found by minimizing $E_{TF}[\rho]$ with the additional requirement
$N = \sum_{\sigma}\int d^{3}\bm{r} \rho_{\sigma}({\bf r})$, which ensures that
the number of particles, $N$, is fixed.

However, DFT guarantees more than just an approximate
description of a many-body system. It actually provides the proof that
there is one-to-one correspondence between the many-body wave function
and the density and therefore various observables can be reconstructed from
the mere knowledge of the density. 
Since the proof, of this theorem (known as Hohenberg-Kohn theorem), is surprisingly
simple, I will present it below.

Suppose, there is a nondegenerate ground state, described by the wave function of $N$ identical spin-$1/2$ particles:
$\Psi(\sigma_{1}\bm{r}_{1},\sigma_{2}\bm{r}_{2},...,\sigma_{N}\bm{r}_{N})$,
which implies that:
\begin{eqnarray}\label{gswf}
\hat{H}\Psi(\sigma_{1}\bm{r}_{1},\sigma_{2}\bm{r}_{2},...,\sigma_{N}\bm{r}_{N})
=E_{gs}(N)\Psi(\sigma_{1}\bm{r}_{1},\sigma_{2}\bm{r}_{2},...,\sigma_{N}\bm{r}_{N}).
\end{eqnarray}
Having $\Psi$ one can construct the one-body density:
\begin{eqnarray}\label{density}
\rho_{\sigma}(\bm{r}) = N_{\sigma} \sum_{\sigma_{2},...,\sigma_{N}=\downarrow,\uparrow}
\int d^{3}\bm{r_{2}}...d^{3}\bm{r_{N}}
|\Psi(\sigma \bm{r},\sigma_{2}\bm{r}_{2},...,\sigma_{N}\bm{r}_{N})|^{2}.
\end{eqnarray}
Let us consider for simplicity the total density distribution assuming 
that the external potential is spin-independent 
$V_{\sigma}^{ext}(\bm{r}) = V^{ext}(\bm{r})$: 
\begin{equation}\label{densitytot}
\rho(\bm{r}) = \rho_{\uparrow}(\bm{r}) + \rho_{\downarrow}(\bm{r}).
\end{equation}

Note, that the following sequence of dependencies hold: 
for each $\hat{V}^{ext}$ one obtains the state $|\Psi [\hat{V}^{ext}]\rangle$,
which is represented by the wave function (\ref{gswf}) and
for each state $|\Psi [\hat{V}^{ext}]\rangle$ one gets 
the density $\rho[\Psi] $ by means of Eqs. (\ref{density},\ref{densitytot}). Consequently, we established the following mappings: 
\begin{equation}
\hat{V}^{ext} \xrightarrow{\text{A}} | \Psi[\hat{V}^{ext}] \rangle \xrightarrow{\text{B}} \rho[\Psi].
\end{equation}

It is easy to show that the map $\text{A}$ is invertible. 
Suppose, we consider two different external potentials:
$\hat{V}_{1}^{ext}=\int d^{3}\bm{r} V_{1}^{ext}(\bm{r})
\sum_{\sigma=\uparrow\downarrow}\hat{\psi}_{\sigma}^\dagger(\bm{r}) \hat{\psi}_{\sigma}(\bm{r})$ and 
$\hat{V}_{2}^{ext}=\int d^{3}\bm{r} V_{2}^{ext}(\bm{r})$ $\sum_{\sigma=\uparrow\downarrow}\hat{\psi}_{\sigma}^\dagger(\bm{r}) \hat{\psi}_{\sigma}(\bm{r})$. Moreover we assume that the two potentials differ by more than a mere additive constant, i.e.,
\begin{eqnarray}\label{potential}
\hat{V}_{1}^{ext}\neq \hat{V}_{2}^{ext} + \mathrm{const}.
\end{eqnarray}
In such a case
\begin{eqnarray}
\hat{H}_{1}\Psi_{1}(\sigma_{1}\bm{r}_{1},\sigma_{2}\bm{r}_{2},...,\sigma_{N}\bm{r}_{N})
&=&(\hat{T}+\hat{V}_{1}^{ext}+\hat{V}^{int})\Psi_{1}(\sigma_{1}\bm{r}_{1},\sigma_{2}\bm{r}_{2},...,\sigma_{N}\bm{r}_{N})= \nonumber \\
&=&E_{1 gs}(N)\Psi_{1}(\sigma_{1}\bm{r}_{1},\sigma_{2}\bm{r}_{2},...,\sigma_{N}\bm{r}_{N}), \\
\hat{H}_{2}\Psi_{2}(\sigma_{1}\bm{r}_{1},\sigma_{2}\bm{r}_{2},...,\sigma_{N}\bm{r}_{N})
&=&(\hat{T}+\hat{V}_{2}^{ext}+\hat{V}^{int})\Psi_{2}(\sigma_{1}\bm{r}_{1},\sigma_{2}\bm{r}_{2},...,\sigma_{N}\bm{r}_{N})= \nonumber \\
&=&E_{2 gs}(N)\Psi_{2}(\sigma_{1}\bm{r}_{1},\sigma_{2}\bm{r}_{2},...,\sigma_{N}\bm{r}_{N}).
\end{eqnarray}
If $|\Psi_{1} \rangle = |\Psi_{2} \rangle$ (apart from the difference due to an arbitrary phase factor, which can be removed), then substracting these two equations one gets:
\begin{eqnarray}
(\hat{V}_{1}^{ext}-\hat{V}_{2}^{ext})\Psi_{1}(\sigma_{1}\bm{r}_{1},\sigma_{2}\bm{r}_{2},...,\sigma_{N}\bm{r}_{N})=\left ( E_{1gs}(N)-E_{2gs}(N) \right ) \Psi_{1}(\sigma_{1}\bm{r}_{1},\sigma_{2}\bm{r}_{2},...,\sigma_{N}\bm{r}_{N}). \nonumber
\end{eqnarray}
However, the action of the external potential operator consists merely of  multiplication by the function describing the external potential:
\begin{eqnarray}
(\hat{V}_{1}^{ext}-\hat{V}_{2}^{ext})\Psi_{1}(\sigma_{1}\bm{r}_{1},\sigma_{2}\bm{r}_{2},...,\sigma_{N}\bm{r}_{N})=\sum_{i=1}^{N}\left ( V_{1}^{ext}(\bm{r}_{i})-V_{2}^{ext}(\bm{r}_{i}) \right ) \Psi_{1}(\sigma_{1}\bm{r}_{1},\sigma_{2}\bm{r}_{2},...,\sigma_{N}\bm{r}_{N}), \nonumber
\end{eqnarray}
which implies that:
\begin{eqnarray}
\sum_{i=1}^{N}\left ( \hat{V}_{1}^{ext}(\bm{r}_{i})-\hat{V}_{2}^{ext}(\bm{r}_{i}) \right )=E_{1gs}(N)-E_{2gs}(N),
\end{eqnarray}
and therefore it clearly violates the assumption (\ref{potential}).
Concluding, the mapping $\text{A}: \hat{V}^{ext}\longrightarrow |\Psi\rangle$ is invertible: 
$\hat{V}^{ext}\longleftrightarrow |\Psi\rangle$.

Let us now consider the second mapping $\text{B}$, and assume that 
$|\Psi_{1} \rangle$ and $|\Psi_{2} \rangle$ ($|\Psi_{1} \rangle \neq \exp(i\alpha) |\Psi_{2} \rangle, \alpha$ - arbitrary real number) describe two ground states of 
$N$-particle
system, corresponding to two external potentials $\hat{V}_{1}^{ext}$ and $\hat{V}_{2}^{ext}$, respectively. It implies that
\begin{eqnarray}\label{2}
E_{1gs}(N) = \langle \Psi_{1}|\hat{H}_{1}|\Psi_{1} \rangle < \langle \Psi_{2}|\hat{H}_{1}|\Psi_{2} \rangle, \\
E_{2gs}(N) = \langle \Psi_{2}|\hat{H}_{2}|\Psi_{2} \rangle < \langle \Psi_{1}|\hat{H}_{2}|\Psi_{1} \rangle,
\end{eqnarray}
where $\hat{H}_{i} = \hat{T} + \hat{V}^{ext}_{i} + \hat{V}^{int}$.
However:
\begin{eqnarray}
\langle \Psi_{2}|\hat{H}_{1}|\Psi_{2} \rangle &=& 
\langle \Psi_{2}|(\hat{T}+\hat{V}^{ext}_{1}+\hat{V}^{int})|\Psi_{2} \rangle = 
\langle \Psi_{2}|(\hat{T}+\hat{V}^{ext}_{2}+\hat{V}^{int}-\hat{V}^{ext}_{2}+\hat{V}^{ext}_{1})|\Psi_{2} \rangle = \nonumber \\
&=&\langle \Psi_{2}|(\hat{H}_{2}-\hat{V}^{ext}_{2}+\hat{V}^{ext}_{1})|\Psi_{2} \rangle 
= \langle \Psi_{2}|\hat{H}_{2}|\Psi_{2} \rangle + \langle \Psi_{2}|(\hat{V}^{ext}_{1}-\hat{V}^{ext}_{2})|\Psi_{2} \rangle = \nonumber \\
= E_{2gs}(N) &+& \sum_{\sigma_{1},...,\sigma_{N}=\downarrow,\uparrow}
\int d^{3}\bm{r_{1}}...d^{3}\bm{r_{N}}
\sum_{i=1}^{N} \left ( V^{ext}_{1}(\bm{r}_{i}) - V^{ext}_{2}(\bm{r}_{i}) \right )
|\Psi_{2}(\sigma_{1}\bm{r}_{1},...,\sigma_{N}\bm{r}_{N})|^{2}= \nonumber \\
&=&E_{2gs}(N) + \int d^{3}\bm{r} \left ( V^{ext}_{1}(\bm{r}) - V^{ext}_{2}(\bm{r})
\right )
\rho_{2}(\bm{r}), \nonumber
\end{eqnarray}
where the last equality follows from Eqs.(\ref{density}),(\ref{densitytot})
and the antisymmetry of the wave function $\Psi_{2}$, which implies
that interchanging variables $(\sigma_{i}, \bm{r}_{i} )\Longleftrightarrow (\sigma_{j}, \bm{r}_{j} )$ does not affect 
the modulus squared $|\Psi(\sigma_{1}\bm{r}_{1},\sigma_{2}\bm{r}_{2},...,\sigma_{N}\bm{r}_{N})|^{2}$.
Therefore one gets:
\begin{eqnarray}\label{1}
\langle \Psi_{2}|\hat{H}_{1}|\Psi_{2} \rangle &=& E_{2gs} + \int d^{3}\bm{r} \left ( V^{ext}_{1}(\bm{r}) - V^{ext}_{2}(\bm{r}) \right )
\rho_{2}(\bm{r}).
\end{eqnarray}
Combining relations (\ref{1}) and (\ref{2}): 
\begin{eqnarray}
E_{1gs} <  E_{2gs} + \int d^{3}\bm{r} \left ( 
V^{ext}_{1}(\bm{r}) - V^{ext}_{2}(\bm{r}) \right )
\rho_{2}(\bm{r}).
\end{eqnarray}
Analogously, considering $\langle \Psi_{1}|\hat{H}_{2}|\Psi_{1} \rangle$ one gets:
\begin{eqnarray}
E_{2gs} <  E_{1gs} + \int d^{3}\bm{r} \left ( V^{ext}_{2}(\bm{r}) - 
V^{ext}_{1}(\bm{r}) \right )
\rho_{1}(\bm{r}).
\end{eqnarray}

If two states $|\Psi_{1}\rangle $ and $|\Psi_{2}\rangle $ generate the same density i.e.,
$\rho_{1}=\rho_{2}$, then adding the above inequalities 
results in
\begin{eqnarray}
E_{1gs} + E_{2gs} <  E_{2gs} + E_{1gs},
\end{eqnarray}
which leads to a clear contradiction. Therefore one can conclude that
the generated densities have to be different: $\rho_{1}(\bm{r})\neq \rho_{2}(\bm{r})$.
Consequently, the mapping $B: |\Psi \rangle \longrightarrow \rho$ is also invertible: 
$\rho \longleftrightarrow |\Psi\rangle$ and 
the quantity:
\begin{eqnarray}
\langle \Psi[\rho]|\hat{H}|\Psi[\rho]\rangle = E[\rho]
\end{eqnarray}
can be treated as a functional of a density $\rho$.

Summarizing, there are three conclusions one can draw from the above considerations:
\begin{itemize}
\item The ground state and the expectation value of any operator $\hat{O}$ can be expressed through
the density $\rho$:  $\langle\Psi[\rho]|\hat{O}|\Psi[\rho]\rangle = O[\rho]$
\item Since both mappings $\text{A}$ and $\text{B}$ are invertible therefore, for a given system, the ground-state density $\rho_{gs}$ is  determined: $\rho_{gs} \longleftrightarrow \hat{V}^{ext} $,
and for a fixed $\hat{V}^{ext}$ the energy is a functional $E_{V^{ext}}[\rho]$, reaches its minimum for $\rho = \rho_{gs}$:
\begin{eqnarray} \label{minimization}
\frac{\delta E_{V^{ext}} }{\delta \rho(\bm{r})} = \mu \Longrightarrow \rho=\rho_{gs},
\end{eqnarray}
where $\mu$ denotes chemical potential ensuring that the number of 
particles is fixed.
Clearly: $E_{V^{ext}}[\rho] > E_{V^{ext}}[\rho_{gs}]$, if $\rho \neq \rho_{gs}$.
\item The invertibility of the mapping $\text{B}: \Psi \longleftrightarrow \rho$ does not depend on $\hat{V}^{ext}$,
which means that there exists a universal functional $F[\rho]$:
\begin{eqnarray} \label{funct}
F[\rho] &=& E_{V^{ext}}[\rho] - \int d^{3}\bm{r} V^{ext}(\bm{r})\rho(\bm{r}) = \nonumber \\
        &=& \langle\Psi[\rho]|\left ( \hat{T} + \hat{V}^{int}
        \right ) |\Psi[\rho]\rangle . 
\end{eqnarray}
The universality means that the functional $F$ is characteristic of the particular system
(depending on the mutual interaction between fermions, their mass, and the kinetic term).
\end{itemize}

\section{Kohn-Sham procedure}

The practical applicability of the Hohenberg-Kohn theorem is related to two key questions:
\begin{enumerate}
\item Can we construct explicitly the functional $F[\rho]$ defined by Eq. (\ref{funct})?
\item Can we solve equations originating from the condition given in Eq.(\ref{minimization})?
\end{enumerate}

Let us set aside the first question for the moment and focus on the second one.
The difficulty in solving Eq.(\ref{minimization}) stems from the following issue: \\
{\em How to perform variation
of the density that is consistent with Pauli principle (and the
constraint: $\int d^{3}\bm{r}\rho(\bm{r}) = N$)? }

The idea behind the Kohn-Sham (KS) procedure is to replace the actual interacting system
of fermions with an equivalent, non-interacting one. Equivalence 
here means
that ground-state densities of both systems are the same.
The Hamiltonian describing the system of non-interacting particles, 
moving in an external potential is given by:
$\hat{H}_{0} = \hat{T} + \hat{V}_{0}$.
In this case, the ground state density can be expressed in the form (I omit
the spin indices):
\begin{eqnarray} \label{ksdensity}
\rho_{0}(\bm{r}) = \sum_{i=1}^{N}|\phi_{i}(\bm{r})|^{2}.
\end{eqnarray}
Consequently, the unique functional is determined solely by the kinetic term:
\begin{equation}
E^{0}_{V_{0}}[\rho_{0}] = \langle\Psi[\rho_{0}]|
\left ( \hat{T} + \hat{V}_{0} \right ) |\Psi[\rho_{0}]\rangle.
\end{equation}

The Kohn-Sham scheme is based on the assumption of {\em v-representability}, which
says that the density of the system of interacting fermions in an external potential $\hat{V}^{ext}$ can be obtained as the density of the form $(\ref{ksdensity})$ 
for some auxiliary potential $\hat{V}_{0}$, which is local~\cite{PhysRev.140.A1133}. 
Therefore, the problem of finding $\rho_{gs}$ is shifted to the problem of finding $\hat{V}_{0}$,
which generates it:
\begin{equation}
\rho_{0}(\bm{r}) = \rho_{gs}(\bm{r})
\end{equation}

Note, that once $\rho_{gs}$ is generated by $\hat{V}_{0}$ for noninteracting system, the uniqueness of such auxiliary external potential is guaranteed by Hohenberg-Kohn theorem. As a consequence one may focus on the minimization of $E^{0}_{V_{0}}[\rho]$ instead 
of $E_{V^{ext}}[\rho]$:
\begin{equation}
\frac{\delta E^{0}_{V_{0}}[\rho]}{\delta \rho} = 
\frac{\delta T[\rho]}{\delta \rho} + V_{0}(\bm{r}) = \mu.
\end{equation}
Comparing the above expression with the variation of $\frac{\delta E_{V^{ext}}[\rho]}{\delta \rho}$:
\begin{equation}
\frac{\delta E_{V^{ext}}[\rho]}{\delta \rho} = 
\frac{\delta T[\rho]}{\delta \rho} + V^{ext}(\bm{r}) +  \frac{\delta V^{int}[\rho]}{\delta \rho}= \mu,
\end{equation}
one gets the formal expression for $V_{0}(\bm{r})$:
\begin{equation}
 V_{0}(\bm{r}) = V^{ext}(\bm{r}) +  \frac{\delta V^{int}[\rho]}{\delta \rho(\bm{r})} =
 V^{ext}(\bm{r}) + \int d^{3}\bm{r'}V^{int}(\bm{r},\bm{r'})\rho(\bm{r'})+
 \frac{\delta V^{corr}[\rho]}{\delta \rho(\bm{r})},
\end{equation}
where in the last equality we have explicitly decomposed the interaction term into Hartree term:
$V^{H}(\bm r) = \int d^{3}\bm{r'}V^{int}(\bm{r},\bm{r'})\rho(\bm{r'})$ and
the so-called {\em correlation} term: $  \frac{\delta V^{corr}[\rho]}{\delta \rho}$ (which contain also the exchange (Fock) term). 

Summarizing: the Kohn-Sham scheme consists of solving self-consistently the following set of equations:
\begin{eqnarray}\label{ks}
\left ( -\frac{\hbar^{2}}{2m}\nabla^{2} + V_{0}(\bm{r}) \right )\phi_{i}{\bm{r}}) &=& 
\epsilon_{i}\phi_{i}{\bm{r}}), \nonumber \\
\rho(\bm {r}) &=& \sum_{i}\theta(\mu - \epsilon_{i}) |\phi_{i}(\bm{r})|^{2},  \\
V_{0}(\bm{r}) &=& V^{ext}(\bm{r}) + V^{H}(\bm r) + 
\frac{\delta V^{corr}[\rho]}{\delta \rho( \bm{r} )}. \nonumber 
\end{eqnarray}
The functions $\phi_{i}$ are called Kohn-Sham (KS) orbitals.

In the set of equations (\ref{ks}), the last relation requires
the knowledge of the functional $F[\rho(\bm{r})]$. 
It may be extremely complicated and highly non-local. 
In problems that are under perturbative control, 
the functional can be formally derived (see e.g., \cite{PhysRevB.32.3876,PhysRevB.15.2884}). However, in the non-perturbative regime - 
such as the unitary Fermi gas~\cite{Zwerger2012_bcsbec}, low energy nuclear systems - one can only adopt a physically motivated approximate functional and subsequently assess its accuracy.
The practical strategy, which is explicitly or implicitly followed in 
many-body systems, like
atomic nuclei or ultracold atomic gases, is the following:
\begin{enumerate}
\item Postulate a simple functional form that captures
the relevant physics with a number of parameters (the fewer the better).
\item Use {\em ab-initio} results (e.g., those provided by Quantum Monte Carlo approach)  
to fix these parameters.
\item Check the accuracy of the functional against known experimental data.
\item If there is a need, improve the form of the functional, using results provided by
other theoretical methods.
\end{enumerate}

The question one faces when applying DFT to a particular system
is how many densities are needed to characterize it.
For example, if we need to account for spin polarization - e.g., triggered by an external magnetic field -
or simply, if we consider a spin-imbalanced system, due to a 
spin-dependent external potential, then it is clear 
that $\rho(\bm{r})$ alone is not sufficient. In such cases, one needs to consider two separate densities $\rho_{\sigma}(\bm{r})$, describing the spin-up and spin-down 
particle distributions, with the additional condition:
\begin{equation}
\int d^{3}\bm{r}\rho_{\sigma}(\bm{r}) = N_{\sigma},
\end{equation}
where $N_{\uparrow}, N_{\downarrow}$ denote the number of spin-up and spin-down particles, respectively.
In general, the need for a particular type of density has to be associated with the external potential,
that couples to it. Hence, for the spin-dependent external potentials 
the necessity of using
both $\rho_{\uparrow}$ and $\rho_{\downarrow}$, as building blocks of the functional is apparent from
the form:
\begin{equation}
F[\rho_{\uparrow},\rho_{\downarrow}] = E_{V^{ext}}[\rho_{\uparrow},\rho_{\downarrow}] - 
\sum_{\sigma=\uparrow\downarrow}\int d^{3}\bm{r} V^{ext}_{\sigma}(\bm{r})\rho_{\sigma}(\bm{r}),
\end{equation}
which is an analogue of Eq. (\ref{funct}).

Another comment must be made regarding the types of quantities we aim to evaluate.
Although the Hohenberg-Kohn theorem tells us that the relevant information
about the wave function is encoded in the density, it is sometimes a highly
non-trivial task to extract it.
Therefore, we divide observables into {\em easy} and {\em hard} categories, depending on how accessible they are within DFT. {\em Easy} observables are those, that are local and can be directly expressed in terms of the local density..
For example, any local operator $\hat{O} = O({\bf r})$ corresponds 
to an observable that is easy to extract, as it simply requires 
to apply directly the prescription:
$\langle\hat{O}\rangle = \int d^{3}\bm{r} O({\bf r})\rho({\bf r})$ using
the density obtained from Eqs. (\ref{ks}).

Suppose, however, that we would like to extract the momentum distribution, or the
S-matrix in a scattering problem. These are {\em hard} observables. 
They can be, 
in principle, extracted but the task is non-trivial. Furthermore, since the functional used is (almost always) not exact, the accuracy of predictions for such observables will generally be much lower than for local ones.


\subsection{Self-bound systems}

The original Hohenberg–Kohn theorem was formulated in the context of systems subject to an external potential, which serves to localize the fermionic density. For electronic systems, which do not form bound states in free space, this framework is natural - the external potential can be interpreted as arising from the Coulomb interaction produced by the distribution of positively charged ions.

However, in self-bound fermionic systems such as atomic nuclei, where the particles form bound states even in the absence of an external field, the situation is fundamentally different. In such systems, the ground-state wave function factorizes (I omit spin degrees of freedom for simplicity) as:
\begin{equation}
\Psi(\bm{r}_{1},\bm{r}_{2},...,\bm{r}_{N}) =
\Phi(\bm{R})\Psi_{int}(\bm{\tilde{r}}_{1},\bm{\tilde{r}}_{2},...,\bm{\tilde{r}}_{N-1}),
\end{equation}
where $\bm{R} = \frac{1}{N}\sum_{i=1}^{N}\bm{r}_{i}$ is the center of mass of the system,
and $\bm{\tilde{r}}_{j} = \bm{r}_{j+1} - \bm{R}_{j}$ with
$\bm{R}_{j}=\frac{1}{j}\sum_{i=1}^{j}\bm{r}_{i}$ are Jacobi coordinates describing
relative motion.

The center-of-mass wave function $\Phi$ is an eigenstate of the total momentum operator and thus produces delocalized density distribution. 
Conversely, the intrinsic component $\Psi_{int}$
encodes the internal structure and yields a localized density in the frame co-moving with the center of mass.

This distinction raises a fundamental question regarding the construction of the energy density functional: Should it be based on the full many-body wave function
$\Psi$ or on its intrinsic part $\Psi_{int}$?
Constructing the energy density functional from $\Psi$
necessitates introducing an external potential to localize the center of mass, leading to a standard Kohn–Sham framework that reproduces the total density~\cite{PhysRevC.77.014311}. However, this procedure inherently entangles internal and center-of-mass degrees of freedom through the KS orbitals, thereby obscuring the physical content of the theory. Since only 
$\Psi_{int}$ captures the intrinsic properties of the self-bound system, artificial  localization of center of mass leads to spurious contributions in the energy
density functional.

A more physically consistent approach is to construct the energy density functional solely from 
$\Psi_{int} $. This requires an external potential that acts within the intrinsic frame. Given that the full Hamiltonian can be decomposed as:
\begin{equation}
\hat{H} = \frac{\bm{P}^{2}}{2mN} + 
\left ( \sum_{i=1}^{N}\frac{\bm{\hat{p}}^{2}_{i}}{2m} - \frac{\bm{P}^{2}}{2mN}
+ \hat{V}^{int} \right ) = \hat{H}_{cm} + \hat{H}_{int},
\end{equation}
and $[\hat{H}_{cm},\hat{H}_{int}]=0$ one can focus on $\hat{H}_{int}$.
The corresponding energy density functional reads:
\begin{equation}
E[\rho_{int} ] = \langle \Psi_{int} |\hat{H}_{int}|\Psi_{int} \rangle +
\int d^{3}\bm{\tilde{r}} \hat{V}^{ext}(\bm{\tilde{r}})\rho_{int}(\bm{\tilde{r}}).
\end{equation}
where
$\rho_{int}(\bm{\tilde{r}}) = 
\int d^{3}\bm{\tilde{r}}_{2}... d^{3}\bm{\tilde{r}}_{N-1} |\Psi_{int}(\bm{\tilde{r}},\bm{\tilde{r}}_{2},...,\bm{\tilde{r}}_{N-1})|^{2}$.
It can be shown, following an adapted version of the original Hohenberg–Kohn argument, that a one-to-one correspondence between $\rho_{int}$ and the external potential 
$V^{ext}(\bm{\tilde{r}})$ also holds in this context~\cite{PhysRevC.75.014306, PhysRevC.76.067302, PhysRevC.80.054314}.

The challenge arises in constructing a Kohn–Sham scheme that reproduces 
$\rho_{int}$ using a system of non-interacting fermions. 
The problem is that the auxiliary potential $\hat{V}_{0} = V_{0}(\bm{\tilde{r}})$ depends on intrinsic (i.e., relative) coordinates, and thus is a many-body operator 
from the laboratory frame perspective. 
It is also easy to realize that the construction of $\rho_{0}$ based
on Eq. (\ref{ksdensity}), which would reproduce $\rho_{int}$ unavoidably 
breaks the translational symmetry.

There is no unique prescription for resolving this difficulty. One possible solution involves introducing a center-of-mass correlation correction into the 
KS potential $V_{0}$. 
This correction accounts for the discrepancy between the interacting intrinsic kinetic energy and the kinetic energy computed from the KS orbitals~\cite{PhysRevC.80.054314}: \\
$\langle\Phi_{int}|
\left ( \sum_{i=1}^{N}\frac{\bm{\hat{p}}^{2}_{i}}{2m} - \frac{\bm{P}^{2}}{2mN} \right )
|\Phi_{int}\rangle - 
\sum_{i=1}^{N}\langle\phi_{i}|\frac{\bm{\hat{p}}_{i}}{2m}|\phi_{i}\rangle $,
where $\phi_{i}$ are single particle orbitals defined by Eq. (\ref{ksdensity}).

In the limit $N\rightarrow \infty$, spontaneous symmetry breaking becomes well-defined, and the center of mass can be treated as a classical variable. This provides a useful reference point for developing systematic approximations 
to the center-of-mass correction~\cite{PhysRevC.76.067302}.

In practice the contribution to the correlation energy related
to the particular symmetry breaking of the intrinsic density, can be
evaluated by projecting out component of the noninteracting
many-body ''wave function'' defined through the orbitals $\phi_{i}$.
This however, makes the simple KS scheme far more complex.
The similar type of problems occur in finite nuclear systems when
other symmetries are broken by the single particle orbitals 
defined through Eq. (\ref{ksdensity}). These include e.g., rotational
symmetry or isospin symmetry.

\section{Superfluidity in Fermi systems}

Superfluidity in Fermi system is associated with the attractive component of interaction, acting between fermions. Although
the mechanisms leading to attractive forces are different
in various systems, the outcome, i.e., the existence of superfluid phase,
can be universally characterized. Namely,
the superfluid phase generates the so-called
off-diagonal long range order (ODLRO)~\cite{RevModPhys.34.694}, which
manifest itself in the behavior of the quantity:
$\langle \hat{\psi}_{ \uparrow}^\dagger\left(\bm{r}_{1}\right)
\hat{\psi}_{\downarrow}^\dagger\left(\bm{r}_{1}\right) 
\hat{\psi}_{\downarrow}        \left(\bm{r}_{2}\right)
\hat{\psi}_{\uparrow}        \left(\bm{r}_{2}\right) \rangle$, 
which is nonzero even if we make $|\bm{r}_{1}-\bm{r}_{2}|$ arbitrarily large:
\begin{equation}
\lim_{|\bm{r}_{1}-\bm{r}_{2}|\to\infty}\langle \hat{\psi}_{ \uparrow}^\dagger\left(\bm{r}_{1}\right)
\hat{\psi}_{\downarrow}^\dagger\left(\bm{r}_{1}\right) 
\hat{\psi}_{\downarrow}        \left(\bm{r}_{2}\right)
\hat{\psi}_{\uparrow}        \left(\bm{r}_{2}\right) \rangle \neq 0
\end{equation}
For a non-interacting Fermi gas, no matter what potential $V_{0}(\bm{r})$ is applied, this condition cannot be fulfilled.
As an example, let us consider the case of a uniform system, in which this quantity can be easily evaluated.
Since the system is translationally invariant, the result 
will explicitly depend on $\bm{r} = \bm{r}_{1}-\bm{r}_{2}$ only:
\begin{eqnarray}
g_{2}(\bm{r}) = \langle \hat{\psi}_{ \uparrow}^\dagger\left( \bm{r}_{1}+\bm{r} \right)
\hat{\psi}_{\downarrow}^\dagger\left(\bm{r}_{1}+\bm{r} \right ) 
\hat{\psi}_{\downarrow}        \left(\bm{r}_{1}\right)
\hat{\psi}_{\uparrow}        \left(\bm{r}_{1}\right) \rangle = \nonumber \\
 \langle \hat{\psi}_{ \uparrow}^\dagger\left(\bm{r}_{1}+\bm{r})\right)
\hat{\psi}_{\uparrow} \left(\bm{r}_{1}\right)\rangle \langle \hat{\psi}_{ \downarrow}^\dagger\left(\bm{r}_{1}+\bm{r})\right)
\hat{\psi}_{\downarrow} \left(\bm{r}_{1}\right)\rangle .
\end{eqnarray}
If we introduce $\hat{\psi}_{\sigma}(\bm{k})$:
\begin{equation}
\hat{\psi}_{\sigma}^{\dagger}(\bm{r}) = 
\frac{1}{(2\pi)^{3}} \int d^{3}\bm{k}\hat{\psi}_{\sigma}^{\dagger}(\bm{k}) e^{-i\bm{k}\cdot\bm{r}},
\end{equation}
one gets:
\begin{eqnarray}
g_{2}(\bm{r}) &=& \left ( \frac{1}{(2\pi)^{3}} \right )^{4}
\int d^{3}\bm{k_{1}}\int d^{3}\bm{k_{2}}\int d^{3}\bm{k_{3}}\int d^{3}\bm{k_{4}}
 \times \nonumber \\
&\times & e^{-i \left ( \bm{k}_{1}\cdot (\bm{r}_{1}+\bm{r}) - \bm{k}_{2}\cdot\bm{r}_{1}
+\bm{k}_{3}\cdot (\bm{r}_{1}+\bm{r}) - \bm{k}_{4}\cdot\bm{r}_{1} \right ) }
\langle\hat{\psi}_{\uparrow}^{\dagger}(\bm{k_{1}})\hat{\psi}_{\uparrow}(\bm{k_{2}})\rangle
\langle\hat{\psi}_{\downarrow}^{\dagger}(\bm{k_{3}})\hat{\psi}_{\downarrow}(\bm{k_{4}})\rangle .
\end{eqnarray}
Since at $T=0$: $\langle\hat{\psi}_{\sigma}^{\dagger}(\bm{k})\hat{\psi}_{\sigma}(\bm{k}')\rangle=
\theta(k_{F} - k )\delta(\bm{k}-\bm{k}')$ and $\mu=\epsilon_{F}=\frac{\hbar^{2}k_{F}^{2}}{2m}$, therefore
\begin{eqnarray}
g_{2}(\bm{r}) &=& \left (\int_{k < k_{F}}  \frac{d^{3}\bm{k}}{(2\pi)^3} e^{-i\bm{k}\cdot\bm{r}} \right )^{2} = 
\left ( \frac{1}{(2\pi)^3} \right )^{2}\left ( 2\pi \int_{0}^{k_{F}} k^{2} dk \int_{-1}^{1} d\cos\theta e^{-i k r \cos\theta} \right )^{2}= \nonumber \\
&=&  \left ( \frac{1}{(2\pi)^3} \right )^{2}\left ( 2\pi \int_{0}^{k_{F}} k^{2} dk \frac{-1}{i k r} \left ( e^{-i k r} - e^{i k r} \right )\right )^{2}=\left ( \frac{1}{(2\pi)^3} \right )^{2}
\left ( \frac{\pi}{r}\int_{0}^{k_{F}} dk k \sin(kr) \right )^{2}= \nonumber \\
&=& \left ( \frac{1}{8\pi^2} \right )^{2}\frac{1}{r^{2}}\left ( \frac{-k_{F}\cos(k_{F} r)}{r} + \frac{\sin(k_{F} r)}{r^{2}} \right )^{2}=
 \left ( \frac{k_{F}^{3}}{8\pi^{2} } \frac{ j_{1}( k_{F} r ) }{k_{F} r} \right )^{2}, \nonumber
\end{eqnarray}
where $j_{1}$ denotes the spherical Bessel function of order one.
It is therefore clear that:
\begin{equation}
\lim_{r\to \infty}g_{2}(r) = 
\lim_{r\to\infty}\left ( \frac{k_{F}^{3}}{8\pi^{2} } \frac{ j_{1}( k_{F} r ) }{k_{F} r} \right )^{2}=0.
\end{equation}
For this reason, in order to capture the phenomenon of superfluidity within the DFT framework, one needs to introduce another type of density. This new density, the so-called {\em anomalous} density, is defined as:
\begin{equation}
\chi_{\uparrow\downarrow} (\bm{r}, \bm{r}') = 
\langle \hat{\psi}_{\downarrow}\left(\bm{r}'\right)\hat{\psi}_{\uparrow}\left(\bm{r}\right)\rangle .
\end{equation}
In such a case, ODLRO will be present as a result of non-vanishing $\chi$:
\begin{equation}
\lim_{|\bm{r}_{1}-\bm{r}_{2}|\to\infty} \langle \hat{\psi}_{ \uparrow}^\dagger\left(\bm{r}_{1}\right)
\hat{\psi}_{\downarrow}^\dagger\left(\bm{r}_{1}\right) 
\hat{\psi}_{\downarrow}        \left(\bm{r}_{2}\right)
\hat{\psi}_{\uparrow}        \left(\bm{r}_{2}\right) \rangle = 
\chi_{\uparrow\downarrow}^{*}(\bm{r}_{1})\chi_{\uparrow\downarrow}(\bm{r}_{2}).
\end{equation}

Hence, one can extend the DFT framework by introducing,
in addition to the external potential $V^{ext}_{\sigma}(\bm{r})$, which couples to the density $\rho_{\sigma}$:
\begin{equation}
E^{ext}[\rho] = \sum_{\sigma=\uparrow\downarrow}\int d^{3}\bm{r} V^{ext}_{\sigma}(\bm{r})\rho_{\sigma}(\bm{r}),
\end{equation}
an external potential $\Delta_{\sigma,\sigma'}(\bm{r},\bm{r}')$, which couples to the anomalous density $\chi$, giving rise to the term:
\begin{eqnarray}
& &E^{ext}[\chi] = \\
&-&\frac{1}{2}\sum_{\sigma,\sigma'=\uparrow\downarrow} \int d^{3}\bm{r}\int d^{3}\bm{r}' 
\left ( \Delta^{ext}_{\sigma,\sigma'}(\bm{r},\bm{r}')\chi_{\sigma,\sigma'}^{*}(\bm{r},\bm{r}') + \Delta^{ext *}_{\sigma,\sigma'}(\bm{r},\bm{r}')\chi_{\sigma,\sigma'}(\bm{r},\bm{r}') \right ). \nonumber
\end{eqnarray}
The minus sign in front of the above expression is merely a matter of convention.

Note, that due to the definition of anomalous density, it has to be antisymmetric:
$\chi_{\sigma, \sigma'}(\bm{r},\bm{r}') = - \chi_{\sigma', \sigma}(\bm{r}',\bm{r})$.
Consequently the potential $\Delta$ has to be antisymmetric as well (or more precisely: the symmetric part of $\Delta$
cancels out in the above expression and does not contribute to $E^{ext}[\chi]$).
Introducing $\chi$ has another important consequence, namely, the particle number is
no longer conserved.
Therefore, we have to introduce explicitly the chemical potential and instead
of $V^{ext}_{\sigma}(\bm{r})$ it is more convenient to consider 
$V^{ext}_{\sigma}(\bm{r})-\mu_{\sigma}$.

Similarly, like in the original Hohenberg-Kohn theorem, one can show the existence of one-to-one correspondence between
densities $\rho$ and $\chi$ and the potentials $V^{ext}_{\sigma}(\bm{r})-\mu_{\sigma}$ and 
$\Delta_{\sigma,\sigma'}(\bm{r},\bm{r}')$~\cite{PhysRevLett.60.2430}.
As a result, the ground state densities can be obtained through the minimization 
of the functional
$E_{(V^{ext}-\mu),\Delta^{ext}}$:
\begin{eqnarray}
\frac{\delta E_{(V^{ext}-\mu),\Delta^{ext}}}{\delta \rho} = 0, \nonumber \\
\frac{\delta E_{(V^{ext}-\mu),\Delta^{ext}}}{\delta \chi} = 0,
\end{eqnarray}
which determine $\rho_{gs}$ and $\chi_{gs}$.
Consequently, it proves the existence of the universal functional describing fermionic superfluid:
\begin{eqnarray}
F[\rho,\chi] &=& E_{(V^{ext}-\mu),\Delta^{ext}} - E^{ext}[\rho] - E^{ext}[\chi] = \\
&=&\langle\Psi[\rho,\chi]|( \hat{T} + \hat{V}^{int} )|\Psi[\rho,\chi]\rangle .
\end{eqnarray}
According to the modified Kohn-Sham prescription,
the densities $\rho$ and $\chi$ can be generated
from a non-interacting Hamiltonian of the form:
\begin{eqnarray} \label{KSsuperfluid}
& &\hat{H}_{0}= 
\sum_{\sigma=\uparrow,\downarrow}\int d^{3}\bm{r}
\hat{\psi}_{\sigma}^\dagger(\bm{r})
\left[-\frac{\hbar^2}{2m}\nabla^2 + V_{0 \sigma}(\bm{r}) - \mu_{\sigma} \right]
\hat{\psi}_\sigma(\bm{r}) - \\
&-& \frac{1}{2}\sum_{\sigma,\sigma'=\uparrow\downarrow}
\int d^{3}\bm{r}\int d^{3}\bm{r}' \left ( \Delta_{0\sigma\sigma'}(\bm{r}, \bm{r}') \hat{\psi}_{ \sigma}^\dagger\left(\bm{r}\right)
\hat{\psi}_{\sigma'}^\dagger\left(\bm{r'}\right)
+ \Delta^{*}_{0\sigma\sigma'}(\bm{r}, \bm{r}')\hat{\psi}_{ \sigma}\left(\bm{r}\right)
\hat{\psi}_{\sigma'}\left(\bm{r'}\right) \right ). \nonumber
\end{eqnarray}
Note, that we have to introduce explicitly the chemical potential in the (grand-canonical) Hamiltonian
(or equivalently replace $\hat{V}_{0\sigma} \rightarrow \hat{V}_{0\sigma} - \mu_{\sigma}$), since 
the last term of the Hamiltonian violates
the particle-number conservation $[\hat{N}, \hat{H}_{0}]\neq 0$, where \\
$\hat{N} = \sum_{\sigma=\uparrow,\downarrow}\int d^{3}\bm{r}  
\hat{\psi}_{\sigma}^\dagger(\bm{r})\hat{\psi}_{\sigma}(\bm{r})$.

The auxiliary Hamiltonian (\ref{KSsuperfluid}) is supposed to generate 
the same ground state densities $\rho$ and $\chi$ as the Hamiltonian describing the interacting system:
\begin{eqnarray}
\hat{H}&=&\hat{T} + \hat{V}^{ext} + \hat{\Delta}^{ext} + \hat{V}^{int} =
\sum_{\sigma=\uparrow,\downarrow}\int d^{3}\bm{r}\;
\hat{\psi}_{\sigma}^\dagger(\bm{r})
\left[-\frac{\hbar^2}{2m}\nabla^2 + V_{\sigma}^{ext}(\bm{r}) - \mu_{\sigma}\right]
\hat{\psi}_\sigma(\bm{r})- \nonumber \\
&-&\frac{1}{2}\sum_{\sigma,\sigma'=\uparrow\downarrow}
\int d^{3}\bm{r}\int d^{3}\bm{r}' \left ( \Delta^{ext}_{\sigma\sigma'}(\bm{r}, \bm{r}') \hat{\psi}_{ \sigma}^\dagger\left(\bm{r}\right)
\hat{\psi}_{\sigma'}^\dagger\left(\bm{r'}\right)
+ \Delta^{ext*}_{\sigma\sigma'}(\bm{r}, \bm{r}')\hat{\psi}_{ \sigma}\left(\bm{r}\right)
\hat{\psi}_{\sigma'}\left(\bm{r'}\right) \right ) \\
&+& \frac{1}{2}\sum_{\sigma,\sigma' =\uparrow,\downarrow}
\int d^{3}\bm{r}\int d^{3}\bm{r'}
\hat{\psi}_{ \sigma}^\dagger\left(\bm{r}\right)
\hat{\psi}_{\sigma'}^\dagger\left(\bm{r'}\right) V^{int}_{\sigma,\sigma'}(\bm{r},\bm{r'})
\hat{\psi}_{\sigma'}        \left(\bm{r'}\right)
\hat{\psi}_{ \sigma}        \left(\bm{r}\right). \nonumber
\end{eqnarray}
This assumption leads to the following expressions for
the external potentials of auxiliary Hamiltonian:
\begin{eqnarray}\label{potdel}
V_{0\sigma}(\bm{r}) = 
 V_{\sigma}^{ext}(\bm{r}) +  \sum_{\sigma'=\uparrow\downarrow}\int d^{3}\bm{r'}V^{int}_{\sigma,\sigma'}(\bm{r},\bm{r'})\rho_{\sigma'}(\bm{r'})+
 \frac{\delta V^{corr}[\rho,\chi]}{\delta \rho_{\sigma}}, \nonumber \\
\Delta_{0\sigma \sigma'}(\bm{r},\bm{r}')=
\Delta^{ext}_{\sigma \sigma'}(\bm{r},\bm{r}') + 
V^{int}_{\sigma,\sigma'}(\bm{r},\bm{r}')\chi_{\sigma,\sigma'}(\bm{r},\bm{r}') 
+ \frac{\delta 
\Delta^{corr}[\rho,\chi]}{\delta \chi} .
\end{eqnarray}
Consequently, the Kohn-Sham procedure can be obtained analogously,
through the requirement of minimization
of $\langle\Psi[\rho,\chi]|\hat{H}_{0}|\Psi[\rho,\chi]\rangle$.
The resulting densities can be expressed as:
\begin{eqnarray}\label{rhochi}
\rho_{\sigma}(\bm{r}) &=& \sum_{n}|v_{n,\sigma}(\bm{r})|^{2}, \\
\chi_{\sigma,\sigma'}(\bm{r},\bm{r}') &=& \sum_{n} v_{n,\sigma}^{*}(\bm{r}) u_{n,\sigma'}(\bm{r}'),
\end{eqnarray}
where $u_{n}$ and $v_{n}$ fulfill the equations:
\begin{eqnarray}\label{hfbks}
\sum_{\sigma'=\uparrow\downarrow}\left ( \hat{h}_{\sigma\sigma'}(\bm{r})u_{n \sigma'}(\bm{r}) + \int d^{3}\bm{r}' \Delta_{0\sigma\sigma'}(\bm{r},\bm{r}')
v_{n \sigma'}(\bm{r}') \right ) = E_{n} u_{n \sigma}(\bm{r}), \\
\sum_{\sigma'=\uparrow\downarrow}\left ( -\hat{h}^{*}_{\sigma\sigma'}(\bm{r})v_{n \sigma'}(\bm{r}) + \int d^{3}\bm{r}' \Delta_{0\sigma\sigma'}^{*}(\bm{r},\bm{r}')
u_{n \sigma'}(\bm{r}') \right ) = E_{n} v_{n \sigma}(n \bm{r}), \nonumber 
\end{eqnarray}
or equivalently
\begin{align}\label{eq:hfb}
\begin{gathered}
\sum_{\sigma'=\uparrow\downarrow}\int d^{3}\bm{r}' \hat{\cal H}_{\sigma\sigma'}(\bm{r},\bm{r'}) 
\begin{pmatrix}
u_{n,\sigma'}(\bm{r}') \\
v_{n,\sigma'}(\bm{r}')
\end{pmatrix}
= E_n
\begin{pmatrix}
u_{n,\sigma}(\bm{r}) \\
v_{n,\sigma}(\bm{r})
\end{pmatrix}, \\
\hat{\cal H} =  
\begin{pmatrix}
\hat{h}_{\sigma\sigma'}(\bm{r},\bm{r}') & \Delta_{0\sigma,\sigma'}(\bm{r},\bm{r}') \\
\Delta^{*}_{0\sigma,\sigma'}(\bm{r},\bm{r}') & -\hat{h}^*_{\sigma\sigma'}(\bm{r},\bm{r}') 
\end{pmatrix},
\end{gathered}
\end{align}
where 
\begin{equation}
\hat{h}_{\sigma\sigma'}(\bm{r},\bm{r}') = \delta_{\sigma\sigma'} \left ( 
-\frac{\hbar^{2}}{2m}\nabla^{2} +  V_{0\sigma}(\bm{r}) - \mu_{\sigma}\right ),
\end{equation}
and $\mu_{\uparrow,\downarrow}$ are chemical potentials for spin-up and spin-down
particles, respectively\footnote{In the above formula we considered for simplicity the operator $\hat{h}$ to be diagonal in spin indices. However, in atomic nuclei, due to spin-orbit potential, the spin-up and spin-down components are coupled. This extension is, however, straightforward.}. 

In fact, the solution of Eqs.(\ref{eq:hfb}) defines the Bogoliubov transformation, which diagonalize ${\cal H}$
and is given by:
\begin{align}
{\cal B} =  
\begin{pmatrix}
U & V^{*} \\
V & U^{*} 
\end{pmatrix},
\end{align}
where $U$ and $V$ are matrices with matrix elements: 
$U_{(\sigma,\bm{r}) (n)} = u_{n \sigma}(\bm{r})$, 
$V_{(\sigma,\bm{r}) (n)} = v_{n \sigma}(\bm{r})$, respectively.
The matrix ${\cal B}$ is unitary:
\begin{eqnarray}
\left ( {\cal B}{\cal B}^{\dagger} \right )_{(\sigma,\bm{r}), (\sigma',\bm{r}')} & =&
\delta_{\sigma\sigma'} \delta(\bm{r}-\bm{r}'), \\
\left ( {\cal B}^{\dagger}{\cal B} \right )_{n,n'} &= &
\delta_{n,n'}, 
\end{eqnarray}
and defines the new creation and annihilation operators: 
($\hat{\alpha}_{n}, \hat{\alpha}^{\dagger}_{n}$), 
which bring $\hat{H}_{0}$ into the diagonal form:
\begin{equation}
\hat{H}_{0} = \sum_{n} E_{n} \hat{\alpha}_{n}^{\dagger}\hat{\alpha}_{n} + const.,
\end{equation}
where
\begin{equation}
\hat{\alpha}_{n} = \sum_{\sigma=\uparrow\downarrow}\int d^{3}\bm{r} 
\left (u_{n\sigma}^{*}( \bm{r} )\hat{\psi}_{\sigma}(\bm{r}) + v_{n\sigma}^{*}( \bm{r} )\hat{\psi}^{\dagger}_{\sigma}(\bm{r}) \right ).
\end{equation}
By taking the hermitian conjugate of the above equation one obtains the expression for $\hat{\alpha}_{n}^{\dagger}$.
The inverse transformation is provided by the hermitian conjugate of the matrix ${\cal B}$ which defines:
\begin{equation}
\hat{\psi}_{\sigma}(\bm{r}) = \sum_{n} \left ( u_{n\sigma}( \bm{r} )\hat{\alpha}_{n} + v_{n\sigma}^{*}( \bm{r} )\hat{\alpha}^{\dagger}_{n} \right ).
\end{equation}
The requirement of unitarity for the Bogoliubov transformation is necessary to preserve the anticommutation
relations:
\begin{equation}
[\hat{\alpha}_{n}, \hat{\alpha}_{n}^{\dagger} ]_{+} =\delta_{n n'}.
\end{equation}

\section{Local $\Delta$ approximation}

The equations (\ref{eq:hfb}) are simplified significantly if the local pairing potential is considered:
\begin{equation}
\Delta_{0\sigma, \sigma'}(\bm{r}, \bm{r}') = \Delta_{0\sigma, -\sigma}(\bm{r}, \bm{r}')\delta(\bm{r} - \bm{r}').
\end{equation}
Note that in this case, due to antisymmetry of $\Delta$, it couples opposite spin states only.
Consequently, Eqs.(\ref{eq:hfb}) become much simpler, as
the integral disappears:
\begin{align}\label{eq:hfbspin}
\begin{gathered}
\hat{\cal H} 
\begin{pmatrix}
u_{n,\uparrow}(\bm{r}) \\
u_{n,\downarrow}(\bm{r}) \\
v_{n,\uparrow}(\bm{r}) \\
v_{n,\downarrow}(\bm{r})
\end{pmatrix}
= E_n
\begin{pmatrix}
u_{n,\uparrow}(\bm{r}) \\
u_{n,\downarrow}(\bm{r}) \\
v_{n,\uparrow}(\bm{r}) \\
v_{n,\downarrow}(\bm{r})
\end{pmatrix}, \\
{\cal H} =  
\begin{pmatrix}
\hat{h}_{\uparrow\uparrow}(\bm{r}) -\mu_{\uparrow}  & \hat{h}_{\uparrow\downarrow}(\bm{r}) & 0 & \Delta(\bm{r}) \\
\hat{h}_{\downarrow\uparrow}(\bm{r}) & \hat{h}_{\downarrow\downarrow}(\bm{r}) - \mu_{\downarrow}& -\Delta(\bm{r}) & 0 \\
0 & -\Delta^*(\bm{r}) &  -\hat{h}^*_{\uparrow\uparrow}(\bm{r}) +\mu_{\uparrow} & -\hat{h}^*_{\uparrow\downarrow}(\bm{r}) \\
\Delta^*(\bm{r}) & 0 & -\hat{h}^*_{\downarrow\uparrow}(\bm{r})& -\hat{h}^*_{\downarrow\downarrow}(\bm{r}) + \mu_{\downarrow}
\end{pmatrix},
\end{gathered}
\end{align}
where $\mu_{\uparrow,\downarrow}$ are chemical potentials for spin-up and spin-down
particles, respectively and $\Delta(\bm{r}) = \Delta_{0\uparrow\downarrow}(\bm{r})$. In this case the second
equation in (\ref{potdel}) takes the form:
\begin{eqnarray} \label{potdellocal}
\Delta_{0\sigma -\sigma}(\bm{r})=
\Delta^{ext}_{\sigma -\sigma}(\bm{r}) + 
V^{int}_{\sigma,-\sigma}(\bm{r})\chi_{\sigma,-\sigma}(\bm{r}) 
+ \frac{\delta 
\Delta^{corr}[\rho,\chi]}{\delta \chi} .
\end{eqnarray}

The locality of the pairing field, despite greatly simplifying the Kohn–Sham equations, introduces a difficulty associated with the divergence of certain quantities.
To understand the origin of this problem, let us consider the case of a uniform, spin-symmetric system 
($\mu_{\uparrow}=\mu_{\downarrow}=\mu$), where $V_{0}^{ext}(\bm{r})=0$ and 
$\Delta$ is simply a constant. 
Then the solution of the above equations can be obtained analytically:
\begin{eqnarray}
u_{\bm{k} \sigma}(\bm{r}) = \sqrt{ \frac{1}{2}\left ( 1 + \frac{ \frac{\hbar^{2}k^{2}}{2m}-\mu}
{\sqrt{\left (\frac{\hbar^{2}k^{2}}{2m}-\mu \right )^{2} + |\Delta|^{2}}}\right ) } e^{ i \bm{k}\cdot\bm{r} }, \\
v_{\bm{k} \sigma}(\bm{r}) = \sqrt{ \frac{1}{2}\left ( 1 - \frac{ \frac{\hbar^{2}k^{2}}{2m}-\mu}
{\sqrt{\left (\frac{\hbar^{2}k^{2}}{2m}-\mu \right )^{2} + |\Delta|^{2}}}\right ) } e^{ i \bm{k}\cdot\bm{r} }.
\end{eqnarray}
Therefore $\chi$ is given by:
\begin{eqnarray}\label{chi}
\chi_{\uparrow \downarrow}(\bm{r}, \bm{r'}) &=& \frac{1}{ (2\pi)^{3} }\int d^{3}\bm{k} 
v^{*}_{\bm{k} \uparrow}(\bm{r}) u_{\bm{k} \downarrow}(\bm{r}') = \\
&=& \frac{1}{ 2(2\pi)^{3} } \int d^{3}\bm{k} \frac{ |\Delta| e^{-i\bm{k} \cdot (\bm{r} - \bm{r}')} }
{ \sqrt{ \left ( \frac{\hbar^{2}k^{2}}{2m}-\mu \right )^{2} + |\Delta|^{2}} }  = \nonumber \\
&=&\frac{1}{(2\pi)^{2}}\frac{|\Delta|}{|\bm{r} - \bm{r}'|}\int_{0}^{\infty}k dk \frac{\sin(k|\bm{r} - \bm{r}'|)}{\sqrt{ \left ( \frac{\hbar^{2}k^{2}}{2m}-\mu \right )^{2} + |\Delta|^{2}} } . \nonumber
\end{eqnarray}
Note that this integral is divergent when $\bm{r}\to\bm{r}'$. For large $k$
the integrand function becomes:
\begin{equation}
k \frac{\sin(k|\bm{r} - \bm{r}'|)}{\sqrt{ \left ( \frac{\hbar^{2}k^{2}}{2m}-\mu \right )^{2} + |\Delta|^{2}} }
\rightarrow \frac{\sin(k|\bm{r}-\bm{r}'|)}{k}
\end{equation}
and therefore, at high momenta, the integral behaves as
\begin{equation}
\int_{k_{cut}}^{\infty}\frac{\sin(k|\bm{r}-\bm{r}'|)}{k} dk = 
\frac{\pi}{2|\bm{r}-\bm{r}'|} - \frac{1}{|\bm{r}-\bm{r}'|} \int_{0}^{k_{cut}|\bm{r}-\bm{r}'|}\frac{\sin x}{x} dx ,
\end{equation}
where we have introduced an arbitrary, positive (and sufficiently large) value of $k_{cut}$.
The second integral on rhs is regular, but the first term diverges at $\bm{r}=\bm{r}'$. Hence it is clear that:
\begin{equation}
\lim_{|\bm{r}-\bm{r}'|\to 0} \chi_{\uparrow \downarrow}(\bm{r}, \bm{r'}) \propto \frac{1}{|\bm{r}-\bm{r}'|}
\end{equation}
In order to get rid of this divergence and keep the pairing field finite one needs to introduce 
the cutoff momentum and
at the same time renormalize the interaction $V^{int}$ in Eq.
(\ref{potdellocal}) \cite{PhysRevLett.88.042504,PhysRevC.65.051305}.
Namely, in the translationally invariant system $V^{int}$ reduces to
a single coupling constant $g$, which has to be replaced by $g_{eff}$:
\begin{eqnarray}
\Delta(\bm{r}) &=& g_{eff}\chi^{reg}_{\uparrow\downarrow}(\bm{r}) \nonumber \\
\chi^{reg}_{\uparrow\downarrow}(\bm{r}) &=& \sum_{E_{n} < E_{cut}}v_{n\uparrow}^{*}(\bm{r})u_{n\downarrow}(\bm{r}) \\
\frac{1}{g_{eff}} &=& \frac{1}{g} + \frac{m k_{cut}}{2\pi^{2}\hbar^{2}}\left (1 - \frac{k_{F}}{2k_{cut}}
\log\left ( \frac{k_{cut}+k_{F}}{k_{cut}-k_{F}}\right ) \right ) \nonumber.
\end{eqnarray}

The above expression for $g_{eff}$ ensures that the pairing gap remains 
finite and independent on $k_{cut}$ (provided $k_{cut}$ is chosen
to be sufficiently large). Note however that the contribution to the 
pairing energy is still cutoff dependent since 
$E_{pair}= -\int d^{3}\bm{r}\Delta^{*}(\bm{r})
\chi_{\uparrow\downarrow}^{reg}(\bm{r})$.
On the other hand, the local pairing field induces also the divergence 
of the kinetic term in the functional. It turns out, however, that evaluating these two quantities, using the same $k_{cut}$, one can reach convergence as a function of the cutoff momentum~\cite{PhysRevC.59.2052, PhysRevLett.88.042504, PhysRevC.65.051305}.
This is due to the fact that the kinetic energy and pairing energy terms behave similarly as functions of $k_{cut}$, and, as a consequence, their contributions cancel out at high momenta.

This particular framework, involving local $\Delta$, is
called {\em Superfluid Local Density Approximation} (SLDA).

\section{Time Dependent Kohn-Sham equations for superfluids}

The theorem relating density to many-body wave functions
in the non-stationary situation has been proved by Runge and Gross~\cite{PhysRevLett.52.997}. It says 
that the densities $\rho(\bm{r})$ and $\rho'(\bm{r})$ evolving from some initial state $|\Psi(t=0)\rangle$,
under the influence of two external potentials $V^{ext}(\bm{r}, t)$ and $V^{'ext}(\bm{r}, t)$
(sufficiently regular, i.e., expandable in Taylor series around $t=0$)
will be different, unless $V^{ext}(\bm{r}, t)- V^{'ext}(\bm{r}, t) = f(t)$, where $f$ is a function of time, only.
Therefore, under this assumption, there is one-to-one mapping between the density $\rho(\bm{r}, t)$
and the potential $V^{ext}(\bm{r}, t)$.
The important component of the proof is the continuity relation $\frac{\partial \rho(\bm{r}, t)}{\partial t}
+\nabla\cdot\bm{j}(\bm{r}, t)=0$ which has to be fulfilled.

Therefore, similarly as in the static case, one can define the fictitious non-interacting system, generating the same time-dependent density distribution, 
as the one describing interacting system of interest. This can be done
by solving the time-dependent Kohn-Sham equations~\cite{ullrich2012time,marques2012fundamentals}:
\begin{eqnarray}
\left ( -\frac{\hbar^{2}}{2m}\nabla^{2} + V_{0}(\bm{r},t) \right )\phi_{i}(\bm{r},t) &=& 
 i\hbar \frac{\partial}{\partial t}\phi_{i}(\bm{r},t), \\
\rho(\bm {r},t) &=& \sum_{i=1}^{N} |\phi_{i}(\bm{r},t)|^{2}, \nonumber \\
V_{0}(\bm{r},t) &=& V_{ext}(\bm{r},t) + V^{H}(\bm r,t) + 
\frac{\delta V^{corr}[\rho, t]}{\delta \rho}. \nonumber 
\end{eqnarray}
This approach can be extended to superfluid systems, leading to the time-dependent version of Eqs.(\ref{hfbks})~\cite{PhysRevLett.73.2915}:
\begin{eqnarray}
\sum_{\sigma'=\uparrow\downarrow}\left ( \hat{h}_{\sigma\sigma'}(\bm{r},t)u_{n \sigma'}(\bm{r},t) + \int d^{3}\bm{r}' \Delta_{0\sigma\sigma'}(\bm{r},\bm{r}',t)
v_{n \sigma'}(\bm{r},t) \right ) = i\hbar \frac{\partial}{\partial t} u_{n \sigma}(\bm{r},t), \\
\sum_{\sigma'=\uparrow\downarrow}\left ( -\hat{h}^{*}_{\sigma\sigma'}(\bm{r},t)v_{n \sigma'}(\bm{r},t) + \int d^{3}\bm{r}' \Delta_{0\sigma\sigma'}^{*}(\bm{r},\bm{r}',t)
u_{n \sigma'}(\bm{r},t) \right ) = i\hbar \frac{\partial}{\partial t} v_{n \sigma}(\bm{r,t}) \nonumber 
\end{eqnarray}
\begin{eqnarray}\label{rhochi1}
\rho_{\sigma}(\bm{r},t) &=& \sum_{n}|v_{n,\sigma}(\bm{r},t)|^{2}, \\
\chi_{\sigma,\sigma',t}(\bm{r},\bm{r}',t) &=& \sum_{n} v_{n,\sigma}^{*}(\bm{r},t) u_{n,\sigma'}(\bm{r}',t),
\end{eqnarray}
\begin{equation} \label{memory}
V_{0\sigma}(\bm{r},t) = 
 V^{ext}_{\sigma}(\bm{r},t) +  \sum_{\sigma'=\uparrow\downarrow}\int d^{3}\bm{r'}V^{int}_{\sigma,\sigma'}(\bm{r},\bm{r'})\rho_{\sigma'}(\bm{r'},t)+
 \frac{\delta V^{corr}[\rho,\chi,t]}{\delta \rho}, 
\end{equation}
\begin{equation} \label{memory1}
\Delta_{0\sigma \sigma'}(\bm{r},\bm{r}',t)=
\Delta^{ext}_{\sigma \sigma'}(\bm{r},\bm{r}',t) + 
V^{int}_{\sigma,\sigma'}(\bm{r},\bm{r}')\chi_{\sigma,\sigma'}(\bm{r},\bm{r}',t) 
+ \frac{\delta 
\Delta^{corr}[\rho,\chi,t]}{\delta \chi}.
\end{equation}

There is, however, one important complication absent in the static case. Portions of the potentials $V_{0}$ and $\Delta_{0}$, arising from the exchange-correlation term, generally retain information about the past evolution of the densities 
$\rho$ and
$\chi$. Namely, in the expressions (\ref{memory},\ref{memory1}) 
the potentials $V_{0}$ and $\Delta_{0}$ at a given
time $t$ depend on densities $\rho(t')$,
$\chi(t')$ for all $0 \le t' \le t$ ($t=0$ being the initial time).
Consequently, the functional governing time evolution is far more intricate than that describing the ground state. These so-called {\em memory effects}, which will appear in $V_{0}$ and $\Delta_{0}$, pose a challenging and still unresolved issue for the application of this framework (see e.g.
Refs \cite{PhysRevB.100.241109,PhysRevA.101.050501} and references therein) .

\section{Adiabatic approximation}

The adiabatic approximation consists of neglecting any dependence on the past included in the correlation terms, i.e., assuming that the following functionals are local in time~\cite{bulgac2012}:
\begin{eqnarray}
\frac{\delta V^{corr}[\rho,\chi,t]}{\delta \rho} = \frac{\delta V^{corr}[\rho(t),\chi(t)]}{\delta \rho(t)}, \\
\frac{\delta \Delta^{corr}[\rho,\chi,t]}{\delta \chi} =
\frac{\delta \Delta^{corr}[\rho(t),\chi(t)]}{\delta \chi(t)}.
\end{eqnarray}
In these expressions, the potentials on the right-hand side depend only on the densities at the given time $t$. This is known as the adiabatic approximation, which becomes exact when the potentials vary so slowly that the particles remain in the instantaneous ground state.

In the adiabatic approximation, and within the local $\Delta$ approximation, the evolution of superfluid system can be obtained from:
\begin{align}\label{eq:hfbspin1}
\begin{gathered}
{\cal H}(t) 
\begin{pmatrix}
\tilde{u}_{n,\uparrow}(\bm{r},t) \\
\tilde{u}_{n,\downarrow}(\bm{r},t) \\
\tilde{v}_{n,\uparrow}(\bm{r},t) \\
\tilde{v}_{n,\downarrow}(\bm{r},t)
\end{pmatrix}
= i\hbar \frac{\partial}{\partial t}
\begin{pmatrix}
\tilde{u}_{n,\uparrow}(\bm{r},t) \\
\tilde{u}_{n,\downarrow}(\bm{r},t) \\
\tilde{v}_{n,\uparrow}(\bm{r},t) \\
\tilde{v}_{n,\downarrow}(\bm{r},t)
\end{pmatrix}, \\
{\cal H}(t) =  
\begin{pmatrix}
h_{\uparrow\uparrow}(\bm{r},t)  & h_{\uparrow\downarrow}(\bm{r},t) & 0 & \tilde{\Delta}_{0}(\bm{r},t) \\
h_{\downarrow\uparrow}(\bm{r},t) & h_{\downarrow\downarrow}(\bm{r},t) & -\tilde{\Delta}_{0}(\bm{r},t) & 0 \\
0 & -\tilde{\Delta}_{0}^*(\bm{r}) &  -h^*_{\uparrow\uparrow}(\bm{r},t)  & -h^*_{\uparrow\downarrow}(\bm{r},t) \\
\tilde{\Delta}_{0}^*(\bm{r}) & 0 & -h^*_{\downarrow\uparrow}(\bm{r},t)& -h^*_{\downarrow\downarrow}(\bm{r},t). 
\end{pmatrix},
\end{gathered}
\end{align}
Note, that the chemical potentials are no longer needed in the evolution. Indeed, 
they can be removed through the following transformation of
amplitudes $u$ and $v$:
\begin{eqnarray}
\tilde{u}_{n,\sigma}(\bm{r},t) = u_{n,\sigma}(\bm{r},t) e^{-i \mu_{\sigma} t/\hbar} \\
\tilde{v}_{n,\sigma}(\bm{r},t) = v_{n,\sigma}(\bm{r},t) e^{ i \mu_{\sigma} t/\hbar}. \nonumber
\end{eqnarray}
Such a change does not affect the density $\rho$ but alters the anomalous density $\chi$:
\begin{equation}
\tilde{\chi}_{\sigma\sigma'}(\bm{r},\bm{r}',t) = \sum_{n}\tilde{v}_{n\sigma}^*(\bm{r},t)\tilde{u}_{n\sigma'}(\bm{r}',t) 
= e^{-i ( \mu_{\sigma}+\mu_{\sigma'}) t }\chi_{\sigma\sigma'}(\bm{r},\bm{r}',t).
\end{equation}
Therefore one can shift the constant term in $h$ to the time-dependent phase factor of $\Delta_{0} \longrightarrow \tilde{\Delta}_{0}=\Delta_{0}e^{-i ( \mu_{\uparrow}+\mu_{\downarrow}) t }$.
Note also, that by construction, the total energy is constant if $\hat{V}^{ext}$ and $\Delta^{ext}$
do not depend on time.
Since the pairing potential $\Delta$
violates particle number conservation, it is not obvious how the particle number evolves in time. In cases where one is interested in the evolution of nuclear systems, keeping the average number of particles constant is of paramount importance. Therefore, one needs to examine the quantity:
\begin{equation}
\frac{d \langle\hat{N}\rangle }{ d t} = \frac{\partial }{\partial t}\sum_{n}\sum_{\sigma=\uparrow\downarrow}
                   \int d^{3}\bm{r}|v_{n\sigma}(\bm{r}, t)|^{2} = \frac{d}{dt} Tr(\rho).
\end{equation}
In order to evaluate this quantity it is useful to consider instead of eq. (\ref{eq:hfbspin1})
the equivalent equations involving explicitly densities $\rho$ and $\chi$.
They fulfill the following equations:
\begin{eqnarray}\label{dens_evol}
i \hbar \frac{d\rho}{dt} &=& [h,\rho] + \Delta_{0}\chi^{\dagger} - \chi\Delta_{0}^{\dagger} \\
i \hbar \frac{d\chi}{dt} &=& h \chi + \chi h^{*} + \Delta_{0} - \Delta_{0}\rho^{*} - \rho\Delta_{0},
\end{eqnarray}
where $\rho$ and $\chi$ are matrices with matrix elements: 
$\rho_{\sigma,\sigma'}(\bm{r},\bm{r}',t) = \langle\hat{\psi}^{\dagger}_{\sigma'}(\bm{r}',t)\hat{\psi}_{\sigma}(\bm{r},t)\rangle$,
$\chi_{\sigma,\sigma'}(\bm{r},\bm{r}',t) = \langle\hat{\psi}_{\sigma'}(\bm{r}',t)\hat{\psi}_{\sigma}(\bm{r},t)\rangle$, as well as
$h$ and $\Delta_{0}$: 
$h_{\sigma,\sigma'}(\bm{r},\bm{r}',t)$, $\Delta_{0 \sigma,\sigma'}(\bm{r},\bm{r}',t)$\footnote{In the Eqs.(\ref{dens_evol}), the terms involving products of two quantities should be understood as matrix multiplications—that is, involving spatial integration and summation over spin indices.}.
Using these relations one can obtain:
\begin{equation}
\frac{d \langle\hat{N}\rangle}{ d t} = 
2 Im \left ( Tr  (\Delta^{ext *}\chi) \right )/\hbar,
\end{equation}
and therefore the average number of particles is conserved as long as the external pairing field: $\Delta^{ext}$ vanishes.

The approach presented in this section—i.e., using the local $\Delta$
approximation and neglecting the memory terms—will be referred to as the {\em
time-dependent Superfluid Local Density Approximation} (TDSLDA).

\section{Applications to nuclear reactions and induced fission}

This part of the lecture will serve more as a review. In the following sections, I will outline three areas where the presented formalism can be applied to the description of superfluid fermions. From this point on, I will use only the SLDA or TDSLDA versions of DFT, as they provide numerically tractable schemes. I will not delve into all the computational details; instead, I encourage readers to consult the referenced materials 
for a more in-depth understanding.

Let us begin with applications in nuclear physics, specifically in the context of nuclear reactions. A more comprehensive discussion of these applications can be found in Ref.~\cite{magierski2019nuclear,pssb.201800592}. Here, I will present a few remarks on typical scenarios where TDDFT proves useful. In nuclear physics, the system is typically assumed to be in its ground state initially, as determined using standard DFT methods. Subsequently, it is acted upon 
by a perturbation that drives the system out of equilibrium. 
The external perturbations can be of various origins: they can be caused by photon absorption, by neutron capture, or
the perturbation arises due to interaction between 
the projectile and the target nucleus (in the case of nuclear collisions). 
It has to be emphasized that the deviations from equilibrium, can be arbitrarily strong. It is one of the most important advantages of
TDDFT, that it can be applied both in the linear-response regime
(where it provides information about excitation energies and spectral properties), 
as well as in the nonlinear regime. In the latter case the external perturbations 
can be strong enough to compete with, or even override the internal interactions 
responsible for the stability of an atomic nucleus.
This is of particular interest for the induced nuclear fission processes, which can be described within TDDFT.

The typical procedure used in the context of TDDFT is the following:
 \begin{itemize}
\item Prepare the initial state, which is usually the ground state (in principle, one
can start from any initial state, but non-stationary initial
configurations are rarely considered and more difficult to obtain in practice). 
This can be achieved by solving static Kohn-Sham equations for an atomic nucleus 
(or nuclei if more than one is involved in the reaction process), to get a
set of ground-state Kohn-Sham orbitals and orbital energies.
\item The time evolution can be obtained by applying certain external field simulating
e.g., the photon absorption, or through generating nonzero velocities of 
nuclei towards each other. 
Then one solves the time-dependent Kohn-Sham equations from the initial time to
the desired final time. The
time propagation of the orbitals generate the time-dependent densities.
\item During time evolution  one may calculate the desired observable(s) as 
functionals of the densities.
\end{itemize}

Pairing correlations are central to understanding various dynamical phenomena in nuclear systems. Among these, low-energy nuclear collisions and induced fission processes stand out as 
the cases where pairing dynamics play a decisive role. In particular, nuclear collisions 
at the energies close to the Coulomb barrier, especially those involving medium-mass or heavy nuclei, provide a unique setting to explore the real-time behavior of superfluid pairing fields. Such studies are also critical in the context of superheavy element synthesis, where precise modeling of the fusion process, taking into account dissipative processes remain an outstanding challenge in nuclear theory~\cite{RevModPhys.72.733, Oganessian_2007}.

Within the TDDFT framework, formulated in the previous sections, pairing correlations are effectively described by a complex pairing field, $\Delta = |\Delta|\exp(i\phi)$, which 
acts as an order parameter for the superfluid phase. The evolution of this field
in the local pairing approximation can be 
naturally treated within TDDFT, which accommodates large-amplitude collective dynamics while remaining computationally tractable, making it suitable for real-time simulations of extended systems (see e.g.,~\cite{pssb.201800592,magierski2019nuclear} and references therein).

\begin{figure}
 \includegraphics[width=0.8\columnwidth]{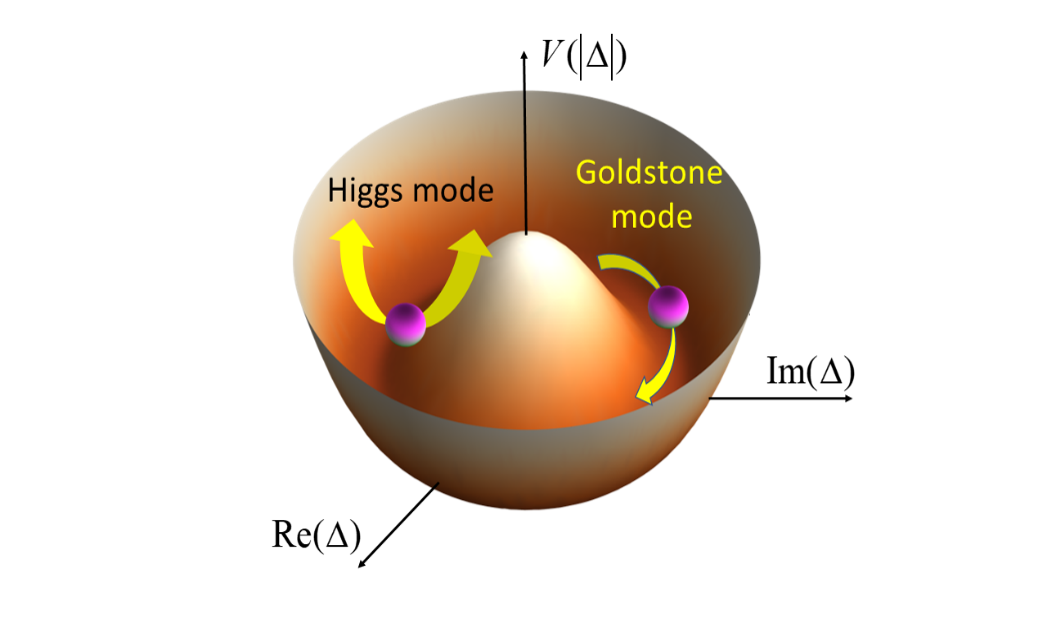}
 \hspace*{1cm}\caption{Schematic illustration of two fundamental excitation modes in a superfluid system.}
 \label{fig:goldstone}
\end{figure}

Two fundamental collective excitation modes emerge from this description: the Goldstone mode and the Higgs mode. These modes correspond, respectively, to phase ($\phi$) and amplitude 
($|\Delta|$) oscillations of the pairing field (see Fig. \ref{fig:goldstone}). The Goldstone mode, in its idealized form, leads to the emergence of Anderson-Bogoliubov phonons — low-energy phase fluctuations well known in bulk superfluids~\cite{Hoinka2017,PhysRevC.99.045801}. However, the finite size of nuclei precludes the formation of long-wavelength excitations. Nonetheless, phase-related dynamics are expected to play a significant role during nuclear collisions.
Two distinct dynamical regimes can be identified in this case. The first pertains to sub-barrier collisions, where the nuclei remain largely intact but a relative phase difference between their respective pairing fields can induce coherent nucleon tunneling. This effect is analogous to the Josephson effect observed in superconducting systems or ultracold gases~\cite{JOSEPHSON1962251, doi:10.1126/science.aac9725} and has been conjectured to occur in
nuclear systems~\cite{1968JETP...26..617G,DIETRICH1970428,DIETRICH1971480,DIETRICH1971201}. 
It has been also predicted that the oscillatory neutron currents between colliding medium-mass nuclei may appear as a nuclear analogue of the AC Josephson effect~\cite{PhysRevC.103.L021601} (see Ref.~\cite{2021PhyOJ..14...27M} for a popular review).

\begin{figure}
 \hspace*{1.0cm}\includegraphics[width=0.8\columnwidth]{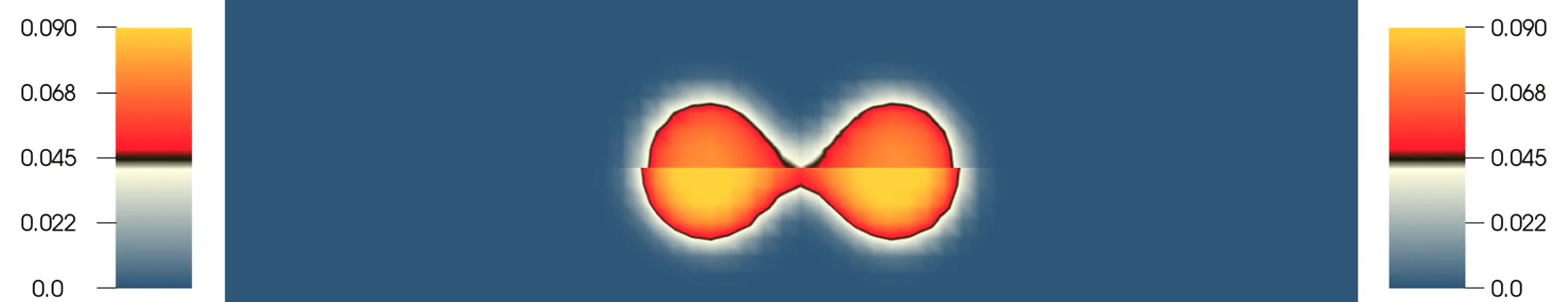}
 \hspace*{1.07cm}\includegraphics[width=0.798\columnwidth]{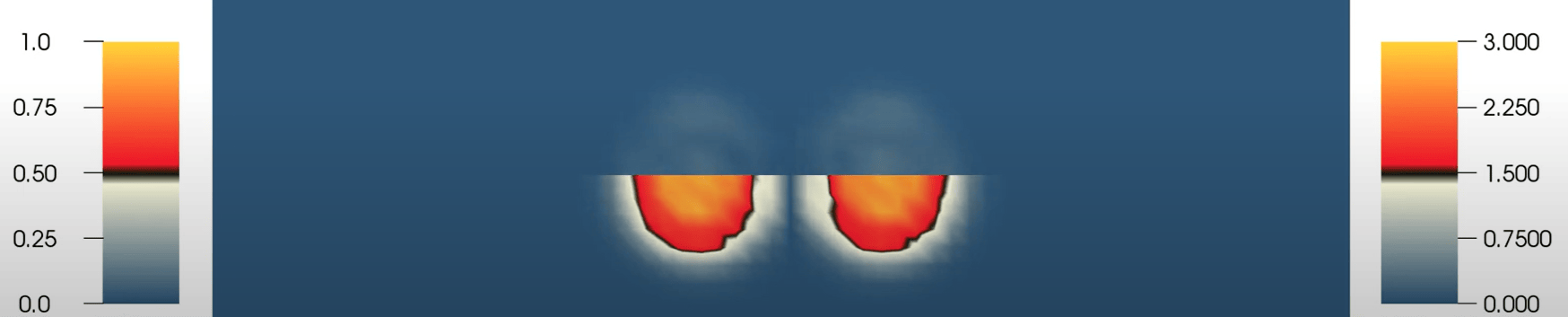}
 \caption{Snapshot from TDSLDA simulation of head-on collision of $^{96}Zr + ^{96}Zr$
 at the center-of-mass energy $182$MeV~\cite{PhysRevC.105.064602}. 
 The upper subfigure shows density distributions
 for protons (upper half) and neutrons (lower half). The lower subfigure shows $|\Delta|$
 for protons (upper half) and neutron (lower half). The solitonic structure appearing
 between colliding nuclei is visible -- $|\Delta|$ is decreased at the 
 contact of two nuclei.
 The phase difference between nuclear pairing fields is equal to $\Delta\phi=\pi$. Legends on the left correspond to protons, whereas on the right - to neutrons.
Magnitudes of $|\Delta|$ are in MeV. Densities are in $fm^{-3}$.}
 \label{fig:collision}
\end{figure}

The second regime is characteristic for the above-barrier collisions. In this case, a solitonic excitation may develop in the neck region between the nuclei~\cite{PhysRevLett.119.042501,PhysRevC.105.064602} (see Fig. \ref{fig:collision}). This soliton-like structure introduces an effective repulsion, thus acting as a transient barrier preventing nuclear capture. Notably, the associated energy penalty depends on the initial phase difference and scales as $\sin^{2}(\Delta\phi/2)$ with the phase difference $\Delta\phi$. Similar phase-driven phenomena have been extensively studied in ultracold atomic gases~\cite{Yefsah2013}, offering valuable insights and analogies for nuclear superfluid dynamics.

These two regimes manifest distinct physical signatures. Sub-barrier dynamics typically enhance nucleon transfer, while above-barrier processes are characterized by the reduced 
capture cross section due to phase-induced solitonic barriers~\cite{PhysRevC.97.044611}. These findings underscore the importance of phase coherence and the non-trivial topological structures of the pairing field in dictating collision outcomes.

Pairing dynamics also emerge as a key factor in the context of induced nuclear fission, where it significantly influences the evolution of the system and the properties of the resulting fragments.
In this process, the compound nucleus, often formed by neutron capture, undergoes large-amplitude collective motion before eventually splitting into two fragments (ternary fission is also possible). This process can be divided into two stages. The first one represents a slow evolution through a multi-dimensional potential energy surface characterized by competing minima and barriers associated with collective degrees of freedom (e.g., quadrupole and octupole deformations). The second stage begins after crossing the outermost barrier and involves a rapid descent toward scission (i.e., the configuration which describes two separate fragments).
The initial phase of fission, spanning timescales of the order or greater than 
$10^{-19}$ seconds, lies beyond the practical reach of TDDFT due to computational constraints. However, the latter phase, leading directly to the scission point, is well-suited for TDDFT. Crucially, it is during this stage that key observables such as total kinetic energy (TKE) and excitation energies of the fission fragments are determined.
TDSLDA simulations have provided new insights into this final phase of fission dynamics~\cite{PhysRevLett.116.122504}. Notably, the collective motion of the nucleus from saddle to scission proceeds at a nearly constant velocity. This behavior is attributed to a highly dissipative process in which the kinetic energy associated with collective deformation is efficiently transferred into intrinsic excitations of the nuclear system~\cite{PhysRevC.100.034615}. As the nucleus elongates and eventually ruptures, the pairing field exhibits both temporal and spatial oscillations, indicative of the excitation of multiple pairing modes~\cite{PhysRevLett.116.122504,PhysRevC.100.034615}.
The strength of the TDDFT approach lies in its ability to provide a self-consistent description of the evolving superfluid field, and allowing for the direct extraction of fragment properties. 
Simulations have shown that TDSLDA can reproduce the experimentally observed TKE values to within 
$1$–$2\%$ accuracy—an impressive result given the complexity of the process~\cite{PhysRevLett.116.122504}. Moreover, it enables the quantification of fragment excitation energies, providing a predictive framework for post-fission fragment evolution.

In summary, time-dependent pairing dynamics reveal as crucial ingredient in both nuclear collisions and fission. From phase-driven transfer and soliton formation in near-barrier collisions to dissipative energy conversion and pairing mode excitation in fission, the TDSLDA formalism emerges as an indispensable tool for probing the non-equilibrium dynamics of superfluid nuclear systems.

An emerging and promising direction involves the integration of TDDFT with projection techniques that restore symmetries broken at the level of the Kohn-Sham (KS) procedure (see section on self-bound systems)~\cite{PhysRevC.100.034612}. This combined approach offers a powerful framework for extracting observables associated with particle number and angular momentum transfer, which are otherwise inaccessible within symmetry-broken mean-field treatments~\cite{PhysRevLett.126.142502,PhysRevC.104.054601, PhysRevC.108.L061602}. 

\section{Applications to neutron star crust modeling}

Neutron stars are the compact remnants resulting from the gravitational collapse of massive stellar cores during core-collapse supernovae. Immediately following their formation, neutron stars possess extreme internal temperatures of about $10^{12}$K. However, they cool 
rapidly—within days—primarily through neutrino emission, reaching temperatures of
$10^{9}$K (see e.g., Ref.\cite{doi:10.1126/science.1090720}).

The interior of a neutron star is characterized by extremely high densities, leading to a highly degenerate quantum state of matter. Under such conditions, various quantum phase transitions are expected to occur, analogous to those observed in condensed matter systems. Of particular interest is the onset of neutron superfluidity. In both the inner crust and the outer core of neutron stars, neutrons are expected to form Cooper pairs, 
akin to electron pairing in conventional superconductors. The theoretical prediction of neutron superfluidity predates the observational confirmation of neutron stars themselves, and its existence is now supported by a growing body of astrophysical evidence and microscopic modeling~\cite{Chamel2017,Sedrakian2019}. The extreme physical conditions inside neutron stars are beyond the reach of laboratory experiments, making direct measurements impossible. As a result, many dynamical aspects of superfluidity—such as vortex dynamics, mutual entrainment between neutrons and protons, and dissipation mechanisms—remain open problems. Insights into these properties must be obtained indirectly via astrophysical observations and interpreted through theoretical modeling.

The inner crust of cold neutron stars exhibits a rich array of nuclear structures resulting from the coexistence of neutron-rich clusters immersed in a superfluid neutron background (see e.g., Ref.~\cite{PhysRevC.105.025806}
and references threrein). In particular, over a specific range of average nucleon number densities, matter organizes into quasispherical nuclear clusters—analogous to impurities—embedded in a neutron superfluid. 
Consequently, there is a compelling need for theoretical paradigms and computational tools tailored specifically to nuclear systems under extreme conditions.
Microscopic modeling plays a central role in the construction of comprehensive global models of superfluid neutron stars. In this context, SLDA and its time-dependent extension TDSLDA have emerged as indispensable tools for probing the static and dynamical properties of dense nuclear matter.

One particularly significant application of TDSLDA is the determination of the vortex pinning force, a quantity of crucial importance for understanding the glitch phenomenon in pulsars. These glitches—sudden increases in rotational frequency—are widely believed to result from the collective unpinning of quantized vortices in the neutron superfluid, as originally proposed by Anderson and Itoh~\cite{ANDERSON1975}.

\begin{figure}
 \hspace*{4.0cm}\includegraphics[width=0.3\columnwidth]{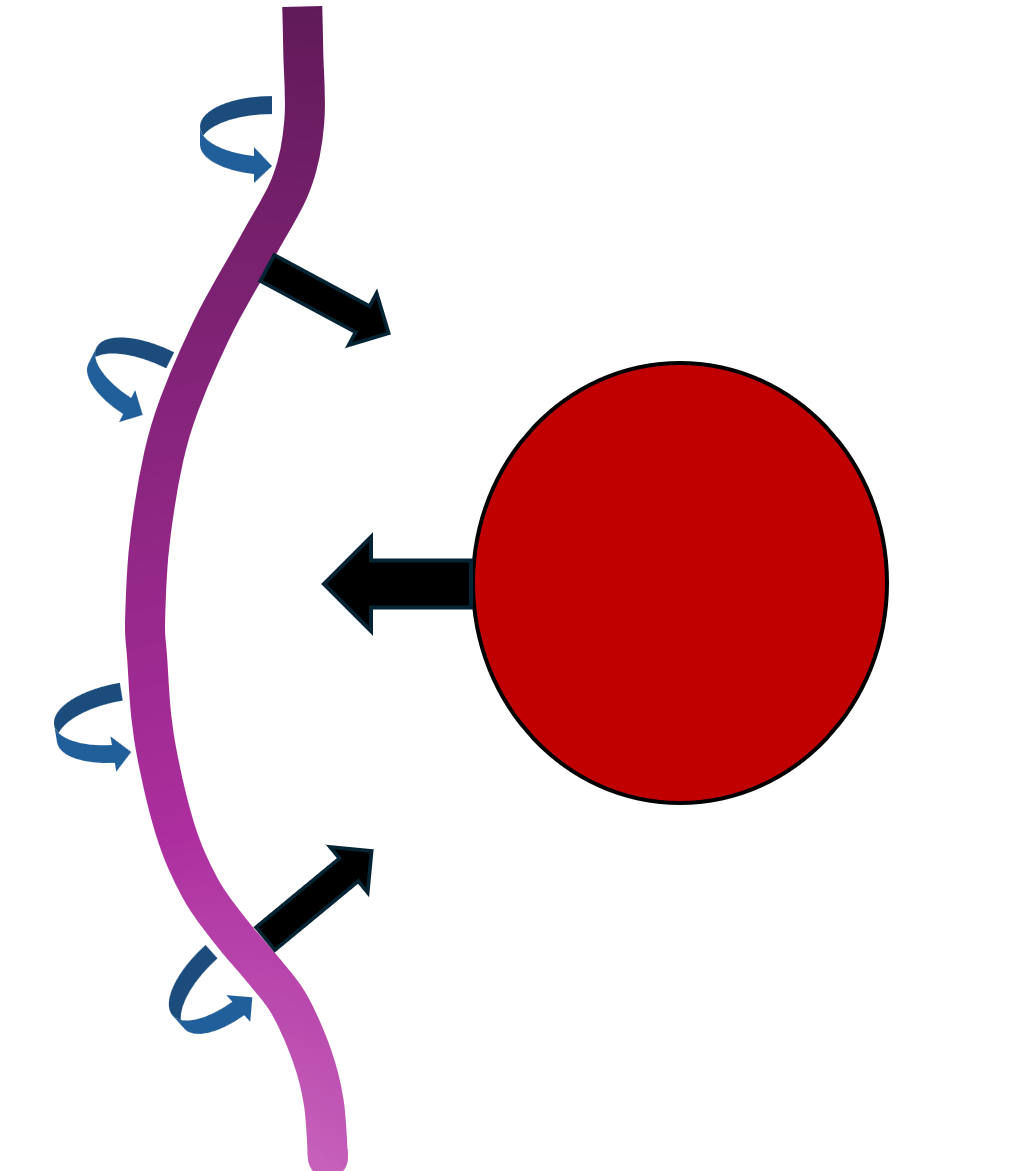}
 \caption{Schematic picture showing vortex line interacting with nuclear impurity. The line
 becomes bent due to the interaction and the nucleus becomes deformed.}
 \label{fig:line}
\end{figure}

Accurately extracting the force between a quantized vortex and nuclear inhomogeneities in the inner crust is therefore vital (see Fig. \ref{fig:line}). 
An application of TDSLDA in this context was presented in 
Ref.~\cite{PhysRevLett.117.232701}, where the interaction between a vortex line and an impurity was quantitatively assessed. This study represents a step toward addressing a broader question: determining the effective equation of motion for a vortex traversing a fermionic superfluid. This subject will be explored in more detail in the following section\footnote{It is worth noting that DFT can also be employed to calculate the pinning energy of a vortex interacting with a nuclear impurity (see, for example, Ref.~\cite{PhysRevC.108.035808} and references therein)}.

Another key quantity characterizing the dynamics of the inner crust is the effective mass of nuclear impurities, which reflects how the surrounding superfluid medium modifies their inertial response. This problem has a long-standing history, dating back to seminal work by Landau, Pekar~\cite{1965478}, and Fr\"ohlich~\cite{Frohlich01} in the context of electrons interacting with lattice vibrations in solids. In nuclear systems, the analog is more complex: the impurity comprises protons and bound neutrons — some of which would be unbound in free space but remain attached due to influence of the surrounding medium. These nuclear clusters are inherently coupled to the neutron superfluid, and their penetrability to neutrons complicates any straightforward treatment of effective mass.
Determining the effective mass of an impurity immersed in a surrounding medium presents significant theoretical challenges, primarily due to the difficulty in accurately evaluating the reversible component of the energy flow between the impurity and the medium. This reversible energy transfer governs the renormalization of the impurity's inertial properties and, hence, its effective mass. In contrast, the irreversible component is associated with dissipative processes.

A variety of approaches have been employed to tackle this problem, often involving simplifying assumptions to render the calculations tractable. Classical hydrodynamic models have been widely used in this context~\cite{Magierski:2003ei,Magierski:2003ej}, providing valuable insights, albeit within a limited framework. 

Time-dependent density functional theory provides a robust framework for simulating real-time dynamics of nuclear impurities within a superfluid environment. This approach enables self-consistent modeling of energy transfer and medium response, even in fully three-dimensional geometries where static approximations become inadequate. In particular, TDSLDA simulations reveal that an impurity initially subjected to a constant external force exhibits linear acceleration—a hallmark of dissipationless motion in a superfluid~\cite{PhysRevX.14.041054}. This is due to the fact that below Landau’s critical velocity, no quasiparticle excitations are possible. 
The first stage of the motion which correspond to dissipationless regime allows to extract the effective mass of the impurity. It can be done simply
by dividing the contant force (acting on protons) by the acceleration of their center of mass.
Calculations of the effective mass have been performed across the entire density range corresponding to the inner crust~\cite{PhysRevX.14.041054}, surprisingly revealing qualitative agreement with results based on irrotational hydrodynamics~\cite{Magierski:2003ei,Magierski:2003ej}. The same framework of TDSLDA can be applied to extract mass parameters associated with any collective nuclear motion~\cite{zdanowicz2025dynamicalschemecomputingmass}.

\section{Applications to ultracold atomic gases}

Ultracold atomic gases represent a rapidly evolving field at the intersection of atomic physics, quantum optics, and condensed matter physics. These systems are composed of neutral atoms cooled to temperatures below $\mu$K, where quantum mechanical effects dominate their behavior. The development of laser cooling and evaporative cooling techniques enabled the experimental realization of such temperatures
(see e.g., Refs. \cite{Ketterle, RevModPhys.80.1215, RevModPhys.80.885}). By trapping atoms in magnetic or optical potentials and manipulating their interactions using Feshbach resonances, they can emulate complex many-body quantum systems. In particular, fermionic gases have become powerful platforms for simulating condensed matter systems, exploring quantum phase transitions, and investigating non-equilibrium dynamics. 

\begin{figure}
 \hspace*{1.0cm}\includegraphics[width=0.8\columnwidth]{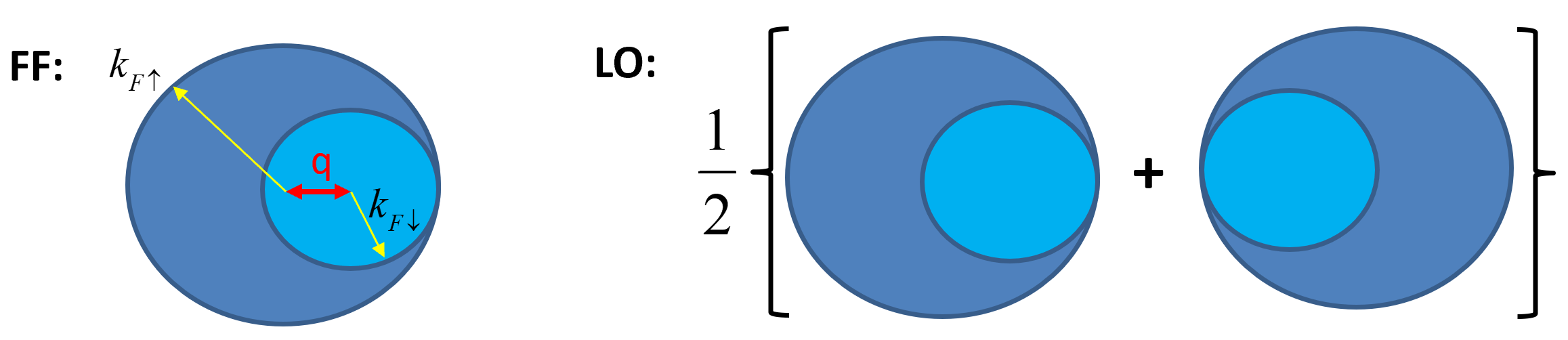}
 \caption{In systems with spin imbalance, the Fermi sphere of the minority spin component (in this case, spin-down fermions with Fermi momentum $k_{F\downarrow}$) is displaced by a momentum vector 
$\bm{q}$ such that it comes into contact with the Fermi surface of the majority (spin-up) component. This momentum mismatch leads to a spatially modulated pairing field. Depending on the structure of this modulation, two distinct phases may emerge:
 Fulde-Ferrell (FF) phase - $|\Delta|\exp(i\bm{q}\cdot\bm{r})$,
 Larkin-Ovchinnikov (LO) phase - $|\Delta|\cos(\bm{q}\cdot\bm{r})$.}
 \label{fig:FFLO}
\end{figure}

One of the interesting features of fermionic gases is the possibility to study the spin-imbalanced
systems, which provide a unique platform for exploring the emergence of exotic superfluid phases and metastable configurations in strongly interacting fermionic systems. The capacity to precisely control spin imbalance has opened up a fertile ground for probing symmetry-broken states that lie beyond conventional BCS-type pairing.
Among the most intriguing theoretical predictions are spatially modulated superfluid phases such as the Fulde-Ferrell-Larkin-Ovchinnikov (FFLO) phase~\cite{PhysRev.135.A550,Larkin:1964wok}, which may serve as a precursor to liquid crystalline order in quantum gases~\cite{PhysRevA.84.023611}, as well as supersolid-like configurations~\cite{PhysRevLett.101.215301} featuring spontaneously broken translational symmetry. 
FFLO phase arise as a consequence of the Fermi momentum mismatch between different spin
components (see Fig. \ref{fig:FFLO}). Rapid advances in trapping and cooling techniques have enabled precise manipulation of spin degrees of freedom, positioning spin polarization as an effective experimental control parameter or "knob"~\cite{doi:10.1126/science.1122318,doi:10.1126/science.1122876,PhysRevLett.97.030401,PhysRevLett.103.170402}). These developments open up the possibility of scanning the full crossover from the Bardeen-Cooper-Schrieffer (BCS) limit, through the strongly interacting unitary regime, to the Bose-Einstein condensate (BEC) side, where a rich landscape of symmetry-breaking phenomena is theoretically anticipated~\cite{SHEEHY20071790,PhysRevLett.103.010404}.

\begin{figure}
 \hspace*{1.0cm}\includegraphics[width=0.8\columnwidth]{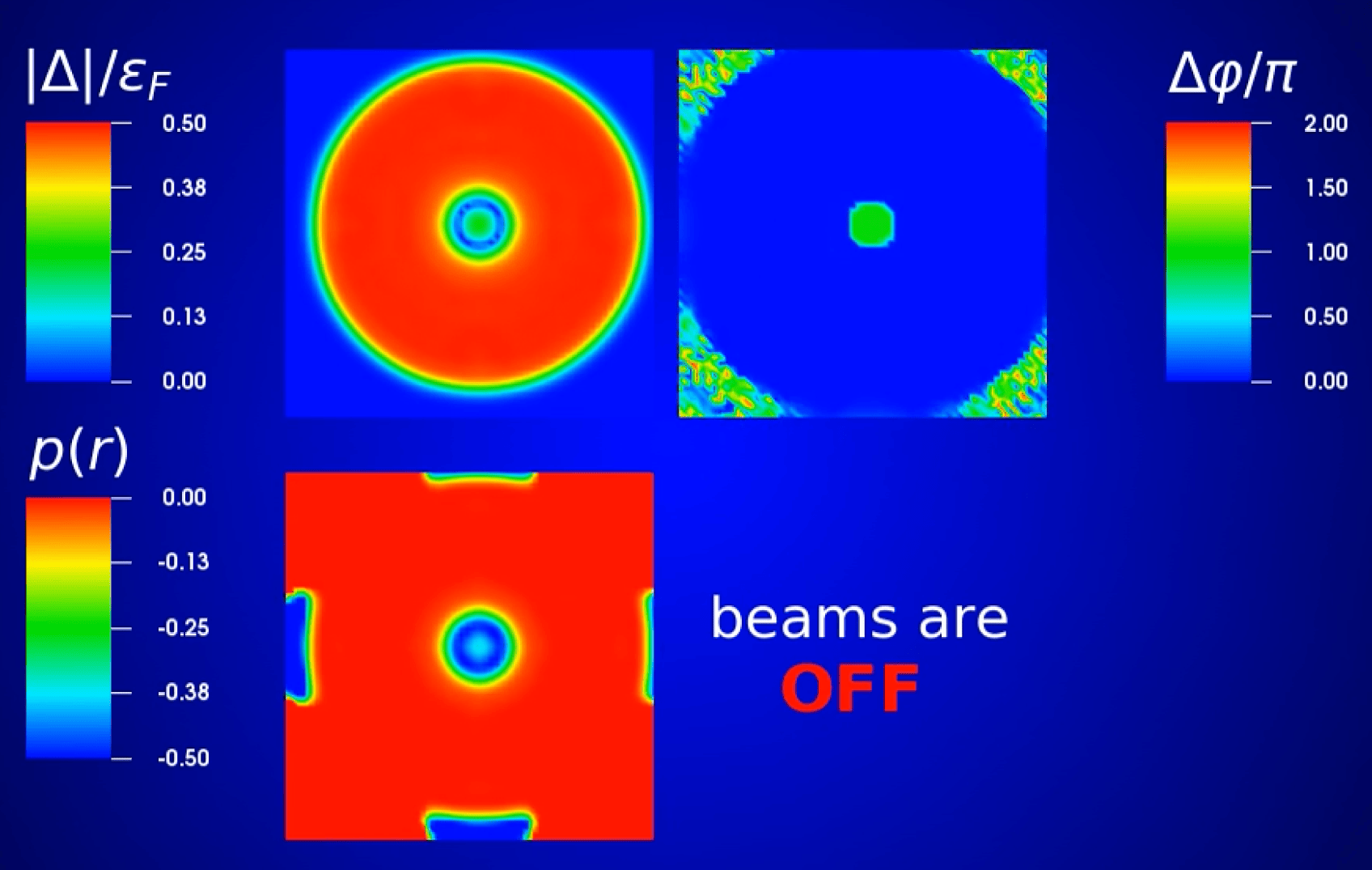}
 \caption{Snapshot from TDSLDA simulations~\cite{PhysRevA.100.033613}, 
 illustrating the formation of a stable structure, referred to as a {\it ferron}, 
 which emerges following the application of a localized, spin-selective potential to an 
 initially spin-unpolarized Fermi gas. This perturbation induces a characteristic nodal 
 structure in the pairing field (upper left panel), closely resembling the spatial modulation 
 found in the Larkin-Ovchinnikov phase. Notably, the sign of the pairing field reverses at 
 the center relative to the surrounding bulk (upper right panel), while the spin polarization 
 becomes concentrated along the nodal line (lower left panel), contributing to the 
 stability of the ferron configuration.}
 \label{fig:ferron}
\end{figure}

A particularly compelling question arises in this context: does a spin-imbalanced Fermi gas support metastable, localized structures capable of storing spin polarization in the absence of global phase separation or bulk FFLO order? Recent work using TDSLDA has addressed this by investigating the dynamical generation of localized spin-polarized droplets—referred to as "ferrons"—in an otherwise unpolarized unitary Fermi gas~\cite{PhysRevA.100.033613,PhysRevA.104.033304}. These structures, reminiscent of finite-size realizations of the Larkin-Ovchinnikov phase, exhibit a distinctive nodal structure in the pairing field and emerge as metastable excitation modes of the superfluid (see Fig. \ref{fig:ferron}).
The metastability of ferrons suggests a novel mechanism for spin localization, raising the possibility that such structures could arise spontaneously during nonequilibrium cooling processes in weakly spin-imbalanced systems. This is particularly relevant in regimes where the spin imbalance is insufficient to stabilize bulk modulated phases like FFLO, yet still energetically favorable for forming localized inhomogeneities.

\begin{figure}
 \hspace*{1.0cm}\includegraphics[width=0.8\columnwidth]{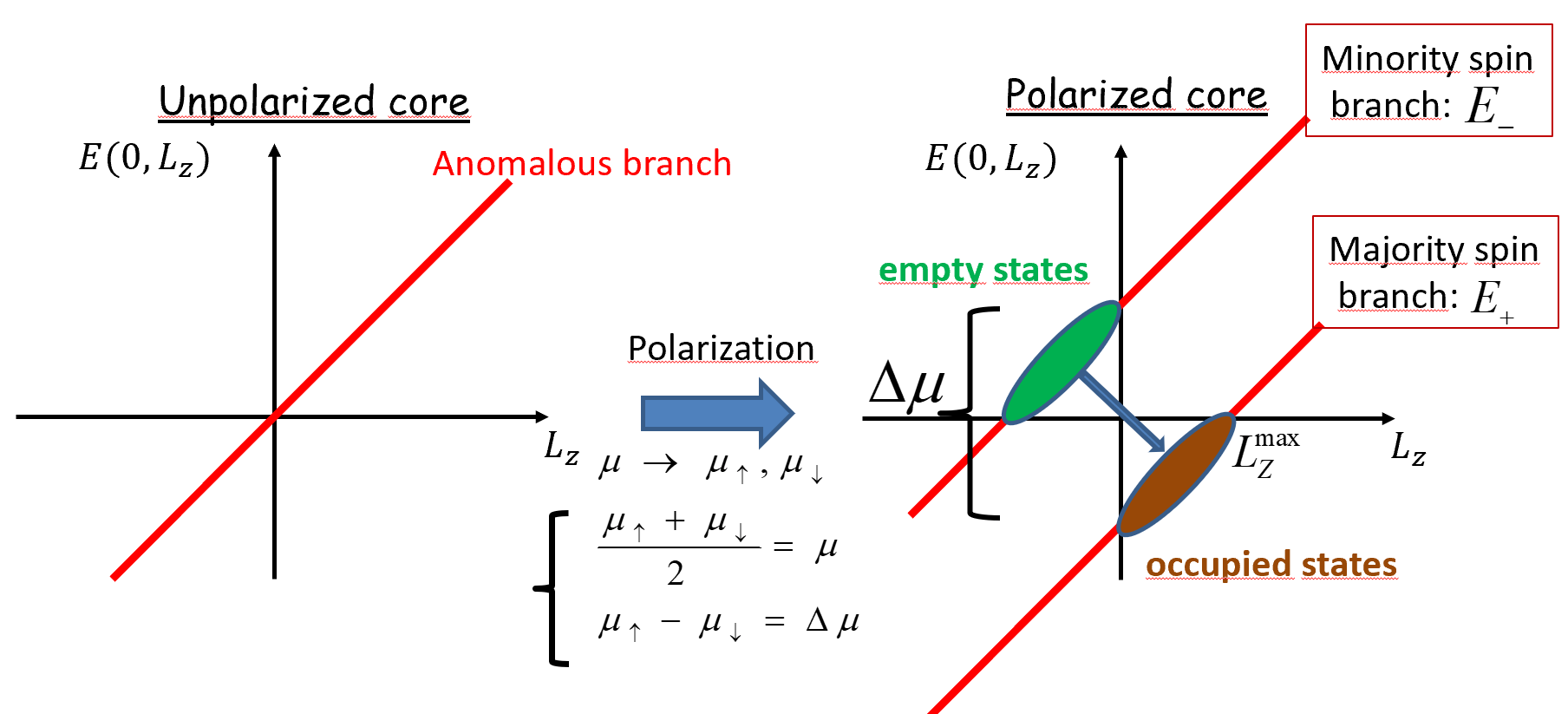}
 \caption{Schematic illustration of quantized states within a vortex core for spin-unpolarized (left) and spin-imbalanced (right) systems. 
 Each branch consists of discrete states. 
 The energy of the states as a function of angular momentum
 along the vortex line (z-axis) forms a straight line for sufficiently
 small $L_{z}$. In the spin-imbalanced case, the branches are shifted relative to each other by the chemical potential difference $\Delta\mu$ between spin-up and spin-down components (see e.g.,
   Ref.~\cite{PhysRevA.106.033322} for the DFT calculations of the states
   in the spin imbalanced vortex core). As a result, the occupation of states within each branch is altered, as illustrated in the right subfigure.}
 \label{fig:vortex}
\end{figure}

Overall, these findings highlight spin-imbalanced ultracold Fermi gases as a valuable platform for exploring metastable states and nonequilibrium dynamics in strongly correlated superfluid systems. They further motivate continued experimental efforts aimed at detecting these elusive structures and charting the spin-polarized phase diagram across the BCS–BEC crossover~\cite{Tüzemen_2023}.

Another intriguing aspect, which was investigated using TDDFT is related to vortex 
dynamics~\cite{Magierski2024}. 
Understanding vortex dynamics lies at the heart of describing superfluid behavior, particularly in relation to dissipative processes that govern the decay of quantum turbulence~\cite{doi:10.1143/JPSJ.77.111006,10.1093/mnras/staa2678,PhysRevA.105.013304,10.1093/pnasnexus/pgae160,PhysRevResearch.6.L042003}. 
As mentioned in the previous section, they are also crucial for our understanding of neutron star glitches~\cite{ANDERSON1975}. 
While the fundamental role of quantized vortices in superfluid hydrodynamics is well established, recent theoretical and numerical advances have emphasized important distinctions between vortices in bosonic and fermionic superfluids—differences that carry significant implications for their dynamical properties.

One critical divergence stems from the intrinsic structure of fermionic vortices, which support a distinct set of low-energy excitations localized within their cores 
(see Fig. \ref{fig:vortex}). In particular, the emergence of a chiral quasiparticle branch—a hallmark of fermionic systems (see e.g., Fig. 8 in Ref.~\cite{PhysRevC.104.055801})—along with the existence of a minigap scale, introduces additional degrees of freedom not present in their bosonic counterparts. These features are expected to contribute non-trivially to the dissipative forces acting on moving vortices, potentially modifying standard phenomenological models derived from bosonic analogues~\cite{Kwon2021,grani2025mutualfrictionvortexhall}.

Another unresolved issue concerns the inertial mass of fermionic vortices. While in bosonic systems the vortex mass is often assumed to be negligible, this simplification may not hold in fermionic systems where core excitations and quasiparticle trapping can result in appreciable mass-like behavior. Determining the mass of a fermionic vortex became recently
possible by applying TDSLDA~\cite{richaud2024dynamicalsignaturevortexmass}

A comprehensive theoretical framework should aim to derive an effective equation of motion for fermionic vortices that incorporates all relevant microscopic contributions. Even in the simplified scenario of two-dimensional (2D) vortex dynamics—where bending of vortex lines and Kelvin wave excitations are neglected—the resulting dynamics are far from trivial. The general form of the vortex equation of motion in such systems must account for inertial, Magnus, drag (mutual friction), all of which demand rigorous microscopic justification from time-dependent density functional theory.

These considerations underscore the pressing need for further investigation into fermio\-nic vortex dynamics. Progress in this direction is not only of theoretical importance but also crucial for interpreting experiments in ultracold atomic gases, neutron stars, and other strongly correlated Fermi systems where superfluidity and quantized vortices play a dominant role.

\section{Summary and outlook}

Density Functional Theory, along with its time-dependent extension, has emerged as a powerful and flexible framework for exploring superfluid phenomena in a wide range of fermionic systems. 
The rapid growth in computational capabilities now permits the real-time evolution of systems comprising tens of thousands of strongly interacting, superfluid fermions, opening new avenues for the study of large-scale quantum dynamics.
In addition to the applications discussed previously, TDSLDA has proven to be a versatile and powerful tool across a broad range of physical systems. Notably, it has been successfully employed to study the generation and decay of pairing Higgs modes in unitary Fermi gases~\cite{Barresi2023}, providing new insights into collective excitations in strongly interacting superfluids.
It has also shed light on dissipation mechanisms in superfluid transport, including the Josephson current~\cite{PhysRevLett.130.023003}, persistent currents
in vortex rings~\cite{xhani2024stabilitypersistentcurrentssuperfluid} and vortex dynamics~\cite{PhysRevLett.130.043001,barresi2025quantumvortexdipoleprobe,grani2025mutualfrictionvortexhall}, where it enabled the identification of underlying microscopic processes responsible for energy loss, both in
spin-unpolarized and spin imbalanced systems.
In the context of spin-imbalanced Fermi systems, the theory has captured complex nonlinear dynamics such as the formation of solitonic cascades~\cite{PhysRevLett.120.253002},
generation of shock waves in the collision of atomic clouds~\cite{PhysRevLett.108.150401} and the emergence of vortex lattices~\cite{PhysRevA.104.053322}, revealing rich structures in the evolution of these out-of-equilibrium states. Furthermore, TDSLDA simulations have provided detailed characterizations of vortex motion and reconnection events, offering quantitative insights into quantum turbulence and superfluid hydrodynamics~\cite{doi:10.1126/science.1201968,PhysRevA.103.L051302,
PhysRevA.91.031602,PhysRevA.103.L051302,Bulgac_2017}.
In nuclear physics, TDSLDA has found important applications as well. It has been utilized to investigate Coulomb excitation processes during nuclear collisions ~\cite{PhysRevLett.114.012701}, and to describe collective excitations such as isovector giant dipole resonances~\cite{PhysRevC.84.051309}, thereby extending its utility to finite, self-bound systems governed by the strong interaction.

This microscopic approach has undergone extensive theoretical validation and has shown strong agreement with experimental observations across both homogeneous and inhomogeneous systems~\cite{Bulgac2012aaa,magierski2019nuclear,pssb.201800592,annurev-nucl-102212-170631}.

State-of-the-art computational implementations, such as the codes developed for
atomic nuclei~\cite{JIN2021108130}, ultracold atomic gases~\cite{WSLDAToolkit}, and neutron stars~\cite{WBSKToolkit}, support both static (SLDA) and dynamic (TDSLDA) simulations of superfluid systems. These software tools, when deployed on leadership-class high-performance computing platforms, are capable of solving millions of coupled, nonlinear, time-dependent, three-dimensional partial differential equations with remarkable efficiency and precision. Notably, the TDSLDA framework enables fully three-dimensional, real-time simulations without imposing any symmetry constraints, offering an unprecedented level of detail in the study of complex superfluid dynamics.

{\bf Acknowledgements}
The author would like to thank Enrico Vigezzi and Francisco Barranco for carefully reading this manuscript and for their numerous valuable comments. The author also gratefully acknowledges Andrzej Makowski, Gabriel Wlazłowski, and Bu\u{g}ra T\"uzemen for their assistance in generating the figures. This work was supported by the Polish National Science Center under Grant No. UMO-2021/43/B/ST2/01191.



\end{document}